\documentclass[11pt,letterpaper]{article}
 
\usepackage{amsmath,amssymb}
\usepackage{array}
\usepackage{cite}

\usepackage{float}
\usepackage{amsmath,amsfonts,graphicx,bm}
\usepackage[usenames]{color}

\usepackage{amsmath,amsfonts,graphicx,bbm,multicol}

\usepackage{subfigure}
\usepackage{titlesec}
\newcommand*{\justifyheading}{\centering}
\titleformat{\chapter}[display]
  {\normalfont\huge\bfseries\justifyheading}{\chaptertitlename\ \thechapter}
  {20pt}{\Huge}
\titleformat{\section}
  {\normalfont\Large\bfseries\justifyheading}{\thesection}{1em}{}
\titleformat{\subsection}
  {\normalfont\Large\bfseries\justifyheading}{\thesubsection}{1em}{}

\newcolumntype{x}[1]{%
{\centering\hspace{0pt}}p{#1}}%

\def\etmiss{E\!\!\!\!\slash_{T}}
\def\ptmiss{p\!\!\!\slash_{T}}

\begin{document}

\baselineskip=18pt

\def\thesection{\Roman{section}} 
\def\thesubsection{\indent\Alph{subsection}}

\newcommand{\eps}{\epsilon}
\newcommand{\pslash}{\!\not\! p}
\newcommand{\I}{\rm 1\kern-.24em l} 
\newcommand{\Tr}{\mathop{\rm Tr}}

\def\wpL{W_L^\prime}
\def\wpR{W_R^\prime}

\def\zpri{Z^\prime}
\def\wpri{W^\prime}

\def\wpLt{\tilde{W}_{L}'}
\def\wpRt{\tilde{W}_{R}'}
\def\mwpL{M_{W_{L}'}}
\def\mwpR{M_{W_{R}'}}

\def\mwpri{M_{W'}}

\def\beq{\begin{equation}}
\def\eeq{\end{equation}}
\def\bea{\begin{eqnarray}}
\def\eea{\end{eqnarray}}
\def\bmat{\begin{pmatrix}}
\def\emat{\end{pmatrix}}
\def\to{\rightarrow}
\newcommand{\chk}[1]{{\bf Check: {\bf #1}}}
\newcommand{\todo}[1]{{\bf To do: {\bf #1}}}
\newcommand{\discuss}[1]{{\bf Discuss: {\it #1}}}
\def\gev{\rm GeV}
\def\tev{\rm TeV}
\def\fbi{\rm fb^{-1}}
\def\lsim{\mathrel{\raise.3ex\hbox{$<$\kern-.75em\lower1ex\hbox{$\sim$}}}}
\def\gsim{\mathrel{\raise.3ex\hbox{$>$\kern-.75em\lower1ex\hbox{$\sim$}}}}
%\def\etmiss{{\overlay{/}{E}}_T}
%\def\ptmiss{{\overlay{/}{p}}_T}
%%%%%%%%  Slash character...
%
\newcommand{ \slashchar }[1]{\setbox0=\hbox{$#1$}   % set a box for #1
   \dimen0=\wd0                                     % and get its size
   \setbox1=\hbox{/} \dimen1=\wd1                   % get size of /
   \ifdim\dimen0>\dimen1                            % #1 is bigger
      \rlap{\hbox to \dimen0{\hfil/\hfil}}          % so center / in box
      #1                                            % and print #1
   \else                                            % / is bigger
      \rlap{\hbox to \dimen1{\hfil$#1$\hfil}}       % so center #1
      /                                             % and print /
   \fi}                                             %
%%EXAMPLE:  $\slashchar{E}$ or $\slashchar{E}_{t}$
\def\ptmiss{\slashchar{p}_{T}}
\def\etmiss{\slashchar{E}_{T}}
\providecommand{\tabularnewline}{\\}

%%%%%%%%%%
%%%%%%%%%%    Title page
%%%%%%%%%%

\thispagestyle{empty}
\vspace{20pt}
\font\cmss=cmss10 \font\cmsss=cmss10 at 7pt

\begin{flushright}
%\today \ \ Ver. 5j \\
%BNL-HET-08/\\
IMSc/2010/08/12 \\
MADPH-10-1559
\end{flushright}

\hfill
\vspace{20pt}

\begin{center}
{\Large \textbf
{
Chiral couplings of $W'$ and top quark polarization at the LHC
}}

\vspace{5pt}

{\large  
Shrihari Gopalakrishna$\, ^{a,b}$,
Tao Han$\, ^{c}$, 
Ian Lewis$\, ^{c}$,
Zong-guo Si$\, ^{d}$,
Yu-Feng Zhou$\, ^{e}$
}

\vspace{10pt}

$^{a}$\textit{The Institute of Mathematical Sciences,
Chennai 600113, India
}
\\
$^{b}$\textit{
School of Physics, University of Melbourne, Victoria 3010, Australia
}
\\
$^{c}$\textit{Department of Physics, University of
Wisconsin, Madison, Wisconsin 53706, USA}
\\
$^{d}$\textit{Department of Physics, Shandong University,
Jinan Shandong 250100, China}
\\
$^{e}$\textit{
Kavli Institute for Theoretical Physics China, 
Key Laboratory of Frontiers in Theoretical Physics, 
Institute of Theoretical Physics, Chinese Academy of Sciences,
Beijing,100190, China
}
\\
\vspace{10pt}
%Dated: 5 Sept 2010
\end{center}

\vspace{15pt}

\begin{center}
\textbf{Abstract}
\end{center}

If a TeV-scale charged gauge boson ($W'$) is discovered at the 
Large Hadron Collider (LHC), it will become imperative
to determine its chiral couplings to standard model (SM) fermions 
in order to learn about the underlying theory containing the $W'$.  
We describe the reconstruction of the $t\, b$ decay mode of the $W'$ at the LHC, 
and identify various kinematic observables 
such as the angular distributions of the top quark and the lepton resulting from top decay
that can be used to disentangle the chiral couplings of the $W'$ to SM fermions. 
We demonstrate by presenting analytical expressions, 
numerical simulations, as well as intuitive illustrations for these observables at  the LHC that among the SM fermions,
the polarized top quark can most directly probe the chirality of such couplings.
\vfill\eject
\noindent

%%%%%%%%%%
%%%%%%%%%%    Main Text
%%%%%%%%%%

%%%%%%%%%%%%%%%%%%%%%%%%%%%%%%%%%%%%%%%%%%%%%%%%%%

\section{INTRODUCTION}
New heavy gauge bosons appear in many  gauge-sector extensions of the Standard Model (SM).
The Large Hadron Collider (LHC) has the potential to discover such vector states \cite{ATLASTDR,CMSTDR}
if their masses are not much larger than a few TeV and their couplings 
to Standard Model states are  not too suppressed.
Once the discovery of a new heavy state becomes established at the LHC,
one would turn next to measuring its properties, such as mass, spin, and couplings.
Kinematical observables such as the invariant-mass or transverse momentum distributions 
of its decay products can be used to extract its mass, 
angular distributions of its production and decay products can be exploited to confirm its spin,
and measured event-rates in various channels can fix combinations of its coupling strengths 
to the  SM fields. 
Another critically important property is the chirality of the new state's couplings to 
SM fermions. Knowing the fermionic chiral couplings can help us glean insight into the 
structure of  the new gauge theory and possibly its gauge-symmetry breaking. 
A prominent example is the left-right symmetric theory with spontaneous parity 
violation \cite{Pati:1974yy}, in which a new heavy charged gauge boson couples to the
SM fermions with right-handed chirality.

If the branching fractions (BR) decaying to the SM leptons are not particularly suppressed, 
the leptonic decays of the new heavy vector state are the much studied  ``golden modes."
The chiral nature of the couplings of a neutral vector boson (generically denoted by $Z'$) can
be probed via the charge forward-backward asymmetry ($A_{FB}$) \cite{Langacker:1984dc}.
This is unavailable at hadron colliders, however, for a charged vector boson (generically denoted by $W'$). 
This follows from an important  fact due to the spin correlation for chiral couplings, that for the
two-body leptonic decays $q\bar q'  \to W^{'-(+)} \to \ell^{-(+)} \nu$, the charged lepton $\ell^{-}$
(antilepton $\ell^{+}$) moves favorably in the direction of 
the initial quark $q$ (antiquark $q'$) direction, 
regardless of the pure left-handed ($L$) or right-handed ($R$) coupling of the $W'$.
However, Ref.~\cite{Cvetic:1993ska} shows that $p p \to \wpri W$ production and $\wpri \to W \ell^+ \ell^-$ decay 
can distinguish the pure left-handed versus pure right-handed cases. 
Although leptonic channels are experimentally clean and will thus be among the first to be observed, 
much information about the heavy vector boson chiral couplings can be learned by analyzing 
the spin correlations of longer decay chains. 
Also, if the leptonic branching fractions are suppressed in a given new-physics scenario
(common in warped extra dimension models \cite{Agashe:2007ki,Agashe:2008jb} for instance),
one will be forced into analyzing more complicated final states including, for instance,
top-quarks.

In this paper, we study in detail the process 
\begin{equation}
p p \to \wpri \to t \bar b \to \ell^+ \nu b \bar b,\qquad (\ell = e,\ \mu)
\end{equation}
and wish to determine the $W'$ chiral couplings by making use of various kinematical variables.
The motivations for this proposal are many:  
First,
the observation of  such a charged  gauge boson $W'$ would not only
unambiguously imply a symmetry beyond a simple $U(1)$ for a $Z'$, 
which could be very common in a variety of scenarios but inconclusive about its origin, but also
most likely predict the existence of a $Z'$ in the same gauge multiplet. 
Second, in some new-physics models,
there are strongly interacting neutral resonances (such as a Kaluza-Klein (KK) gluon) fairly close in mass 
to the $Z'$s. These resonances would decay to $t \bar t$ with much larger rates, masking the 
electroweak neutral gauge boson and causing it to be
difficult to stand out in the $t\bar t$ channel \cite{Djouadi:2007eg}.
The mode $\wpri \to t b$ final state, on the other hand, would be in principle free from this problem. 
Third,
a $W'$ typically leads to a sizable production rate and larger decay branching
fraction to the top, with a slightly more favorable phase space. 
(See also Ref.~\cite{Boos:2006xe} for the $W'\to t b$ cross section.)
Finally, as we will demonstrate later, the angular distribution of the lepton reconstructed in the
top-quark rest frame serves as a diagnostic of the top spin which in turn contains 
information on the chirality of the $\wpri$ coupling. 
This point above was first considered in \cite{Frere:1990qm}. 
We stress that this feature is unique for the top-quark 
final state, and unavailable to any other $W'$ decay modes to SM particles. 
We will utilize techniques that determine the quark direction on a statistical basis.
We will discuss issues in the reconstruction of the top-quark and how detector resolution
and SM backgrounds affect the determination of the $\wpri$ chirality.  
The discussion will be kept model-independent,  as long as 
the $\wpri$ branching fraction to $t \bar b$ is not too small 
and that the top-quark decays are essentially as in the SM. 
This assumption is not particularly restrictive since  it has been shown~\cite{InsenWtb} 
that the lepton angular distribution 
from top decay is insensitive to the presence of small anomalous $t b W$ couplings. 

The paper is organized as follows: 
In Sec.~\ref{ThFrame.SEC} we specify a 
model-independent theoretical framework defining the $W'$ chiral couplings to SM fields 
and  list the constraints on $W'$ bosons from current experiments.
In Sec.~\ref{WpLHC.SEC}, we discuss techniques to reconstruct the event 
for the $W' \rightarrow t b$ decay mode at the LHC.
In Sec.~\ref{WpriAC.SEC} we present a detailed exposition of ways to determine the
chiral couplings of the $W'$ at the LHC using many complementary measurements, namely: 
(i) the angular distribution of the top-quark;  
(ii) the angular distribution of the lepton in the top rest frame with respect to the top moving direction; 
(iii) the lepton transverse momentum distribution;
(iv) the angular distribution of the lepton in the $W$ rest frame with respect to the $W$ moving direction. 
We end with discussions and conclusions in Sec.~\ref{DiscConcl.SEC}. 
The analytical expression for the angular distribution of the lepton in the top-quark rest-frame
is fully worked out in Appendix~\ref{LepAngDist.APP}, and the technical details of evaluating 
a contour integral encountered there is given in Appendix~\ref{ContInt.APP}. 

%%%%%%%%%%%%%%%%%%%%%%%%%%%%%%%%%%%%%%%%%%%%%%%%%%%%%%%%%%%%%%%%%%%%%%%%%%%%%%%%
\section{THEORETICAL FRAMEWORK}
\label{ThFrame.SEC}
%\subsection{}
Many beyond the SM theories contain new heavy charged vector bosons that couple to the SM.
For example, a left-right symmetric theory \cite{Pati:1974yy}
has a charged gauge boson, $W_L$,  associated with the SM $SU(2)_L$ gauge group 
along with a new (heavy) charged gauge boson, $W_R'$, associated with the $SU(2)_R$ gauge group.
Little Higgs models often introduce an enlarged gauge symmetry that includes new 
gauge bosons \cite{ArkaniHamed:2002qy}. 
Other examples are the extradimensional theories in which KK excitations of bulk 
gauge bosons appear as heavy vector states, such as in the universal extra dimensions \cite{Appelquist:2000nn},
in Higgsless models \cite{Csaki:2003zu}, 
and in warped-space models in which the $SU(2)_R$ is gauged~\cite{Agashe:2003zs} 
along with the $SU(2)_L$ where
the KK gauge bosons may have highly suppressed couplings to the SM light fermions. 
For the purpose of this work we will not specialize to a specific theory.
Instead, we parametrize the chiral coupling in a generic form with a minimal set of parameters.

%%%%%%%%%%%%%%%%%%%%%%%%%%%%%%%%%%%%%%%%%%%%%%%%%%%%%%%%%%%%%%%%%%%%%%%%%
\subsection{$W'$ Coupling to SM Fermions}

We wish to keep this analysis model independent and will 
parametrize the $\wpri$ coupling to SM fermions ($\psi$) using the general form
\beq
{\cal L} = \frac{ g_{2} }{\sqrt{2}}\bar{\psi_u^i}\gamma_{\mu}
\sum_{\tau=L,R}g_{\tau}V_{\tau}^{\prime ij}P_{\tau}\psi_d^j{{W'^{+}_{\tau}}^{\mu}}
 + {\rm h.c.} \ ,
\label{Wpff.EQ}
\eeq
where $(\psi_u, \psi_d)$ are mass-eigenstate fermions and form a $SU(2)_L$ or $SU(2)_R$ doublet; 
$i,j$ the generation indices; $V^\prime$ a unitary matrix 
representing the fermion flavor mixing; and
$P_{L,R}=(1\mp \gamma_5)/2$ the left- and right-chirality projection operators.
In our work, we will thus restrict ourselves to renormalizable operators and consider only
flavor-diagonal effects, but otherwise will not specialize to any particular model of new physics.

For the sake of convenient comparison with the SM results, we have written the overall coupling strength 
in terms of
the $SU(2)$ coupling $g_2$ in the SM. 
Some representative choices are 
\bea
g_{L,R}^{ij}\equiv g_{L,R} V^{\prime ij}_{L,R}\ , \quad
\left\{  
\begin{array}{ll}
g_{L}=1,\quad g_{R}=0, \quad   {\rm  pure\  left-handed\  gauge\ boson }\ W'_{L}, & \\
g_{L}=0,\quad g_{R}=1, \quad   {\rm  pure\  right-handed\  gauge\ boson}\ W'_{R}, & \\
g_{L} = g_{R}=1, \quad   {\rm   left-right\  symmetric\ gauge\ boson}. & \\
\end{array}
\right.
\eea

Although for illustrations and plots we consider the same-chirality couplings of the $\wpri$ to the 
$q q^\prime$ and to the $t b$, i.e. either both left-handed or both right-handed as above,
our analytical expressions are given in full generality.  
In some cases there could be more than one $W'$ state with different $g_L$ and $g_R$.
These states could contribute to the same final state and their
amplitudes would have to be added coherently, depending on their mass degeneracy. 
Indeed, if some of these states are quite close in mass, there could arise interesting structure in the
squared amplitude which could be reflected in the angular correlations. 
In this work we will explore only a single $W'$ as the lower-lying state beyond the SM.
The techniques we develop here could easily be extended to cases with multiple $W'$ states.

In many models of new physics, the $W' W Z$ coupling arises after electroweak symmetry breaking
and is typically suppressed by a heavy-light mixing angle 
({\it e.g.}~Refs.~\cite{Agashe:2007ki,Agashe:2008jb,Csaki:2003zu}). 
This coupling can be written as
\beq
g_{W' W Z} \sim g^{SM}_{WWZ} \left( \frac{M_{W}}{M_{W'}} \right)^2  ,
\label{WpWZcoup.EQ}
\eeq
$M_{W}$ and $M_{W'}$ are the masses of the SM gauge boson  $W^{\pm}$ and $\wpri$, respectively.
The mixing angle suppression may potentially be overcome due to an enhancement of the heavy $\wpri$ 
decay into longitudinally polarized $W$ and $Z$.
For a $\wpri$ coupling to fermions with SM strength and the $W' W Z$ coupling as in Eq.~(\ref{WpWZcoup.EQ}), we estimate that the BR of a $1$~TeV $\wpri$ to SM $W$ and $Z$ is approximately $1.3\%$.  Similarly, the BR of a $1$~TeV $\wpri$ to SM $W$ and Higgs boson is about $2\%$.
We will therefore not consider the decay of a heavy $\wpri$ into SM gauge bosons, 
but in a particular new-physics scenario the BR into these modes could be sizable\footnote{For
example in the Randall-Sundrum model in Ref.~\cite{Agashe:2007ki} this coupling is enhanced for longitudinal 
modes by the volume factor and turns out to be significant.}.

In this work, we will mainly focus on heavy charged electroweak states decaying to heavy quarks
$W' \to t b.$ As will be seen later, there is a major advantage to consider the heavy fermion final state
to determine the $W'$ chiral coupling due to the correlation via the top-quark polarization.

%%%%%%%%%%%%%%%%%%%%%%%
\subsection{Constraints on the $W'$ Mass}
Constraints on a new gauge boson of $\zpri$ or $\wpri$ come mainly from direct collider searches, 
indirect bounds from the precision electroweak observables, and flavor-changing-neutral-current processes. 
We now summarize the ones relevant to our interests.

\noindent
(1)
Direct searches at the Tevatron:
\begin{enumerate}
\renewcommand{\theenumi}{(\roman{enumi})}
\renewcommand{\labelenumi}{\theenumi}
\item Assuming no $W-W'$ mixing, SM-type couplings and the decay channel $W'\to WZ$ is 
suppressed, a search in the leptonic decay mode $e\nu$ by the CDF collaboration set
a bound of  $\mwpri>788$~TeV for manifest left-right symmetric model~\cite{Abulencia:2006kh}.
The D0 collaboration recently obtained a lower bound of $\mwpri>1$~TeV \cite{:2007bs}. 

\item With right-handed couplings, the absence of the $t \bar{b}$ signal at CDF gives 
$\mwpri>800$~GeV including the $W'$ leptonic decays,  and $\mwpri>825$~GeV if leptonic
decays are forbidden~\cite{Aaltonen:2009qu}.
\end{enumerate}
(2) Precision electroweak bounds
\begin{enumerate}
\renewcommand{\theenumi}{(\roman{enumi})}
\renewcommand{\labelenumi}{\theenumi}
\item A recent global fit to various $SU(2)_1\times SU(2)_2\times U(1)_X$
  models shows that if the gauge couplings are taken as free parameters, the
  bound on $\mwpri$ is typically below 1~TeV while the mass of $Z'$ is above
  $1.5$ TeV \cite{Hsieh:2010zr}.

\item Bounds resulting from a fit to precision electroweak observables for
  generic $Z'$ are found in Ref.~\cite{Cacciapaglia:2006pk} to be $
  M_{Z'}/g_{Z'} \gtrsim 2.7 - 6.7$~TeV. 
  Similar bounds can be applied for the
  $W'$ since in typical models the $W'$ mass is related to the $Z'$ mass by an
  $O(1)$ factor, although it is conceivable that in certain models the
  splitting could be larger.
  For a related discussion on such bounds, see also Ref.~\cite{Cvetic:1993ska} and references therein. 
\end{enumerate}
(3) Low energy bounds
\begin{enumerate}
\renewcommand{\theenumi}{(\roman{enumi})}
\renewcommand{\labelenumi}{\theenumi}
\item Global analysis from muon decays, electron-hadron, neutrino-hadron and 
neutrino-electron interactions in manifest left-right models give $\mwpri>715-875$~GeV, depending on the $W_L-W_R$ mixing \cite{Czakon:1999ga}.

\item Bounds from flavor-changing-neutral-current processes in the minimal left-right symmetric model can be as severe 
as $\mwpri>2$~TeV~\cite{LR-bound}, but quite model dependent.  
With a certain form of the right-handed quark mixing (Cabibbo-Kobayashi-Maskawa quark-mixing), the bound can be relaxed 
to  about $300$~GeV~\cite{Langacker:1989xa}.  For left-right models with an extended Higgs sector, 
the bound can also be relaxed to about  1~TeV depending on the structure of the
Higgs sector ~\cite{Wu:2007kt,Guadagnoli:2010sd}.

\end{enumerate}

We thus will primarily study  the case $\mwpri = 1$ TeV for illustration, and will comment on
the situation for a heavier one. 

%%%%%%%%%%%%%%%%%%%%%%%%%%%%%%%%%%%%%%%%%%%%%%%%%%%%%%%%%%%%%%%%%%%%%%%%%%%%%%%%%%%%%%%%%%%%%
\section{SIGNAL FOR $\wpri \to t b$ AT THE LHC}
\label{WpLHC.SEC}

\begin{figure}[tb]
\centering
\includegraphics[width=0.5\textwidth,angle=0]{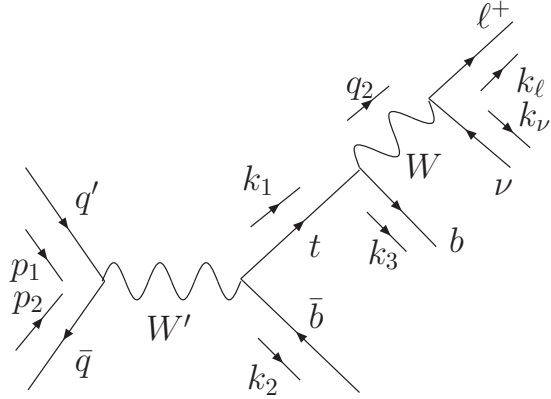}
\caption{
The partonic level process for a heavy ${\wpri}^+$ production in hadronic collisions. 
%\todo{Add $p_2$ label.}
}
\label{qq2Wp2tb.FIG}
\end{figure}

%\subsubsection{$W'\rightarrow t\overline{b}$}
At the LHC, the dominant parton-level subprocess for a heavy $W'$ production is depicted in 
Fig.~\ref{qq2Wp2tb.FIG}, as labeled with the corresponding momenta
\beq
q'(p_{1})\  \bar q(p_{2}) \to W'^{+} \to t(k_{1})\ \bar b(k_{2})  \to W^{+}(q_{2})\ b(k_{3}) \ \bar b(k_{2}) \to
\ell^{+}(k_{\ell})\  \nu(k_{\nu}) \  \ b(k_{3}) \ \bar b(k_{2}),
\eeq
plus its conjugate process of $W'^-$ production with a smaller rate. 
We wish to identify the signal events with 
a very energetic charged lepton, two high-energy $b$-quark jets, and large missing energy
from the undetected neutrino.

%
%%%%%%%%%%%%%%%%%%%%%%%%%%%%%%%%%%%%%%%%%%%
\subsection{$\wpri$ Production and Decay}
\begin{figure}[tb]
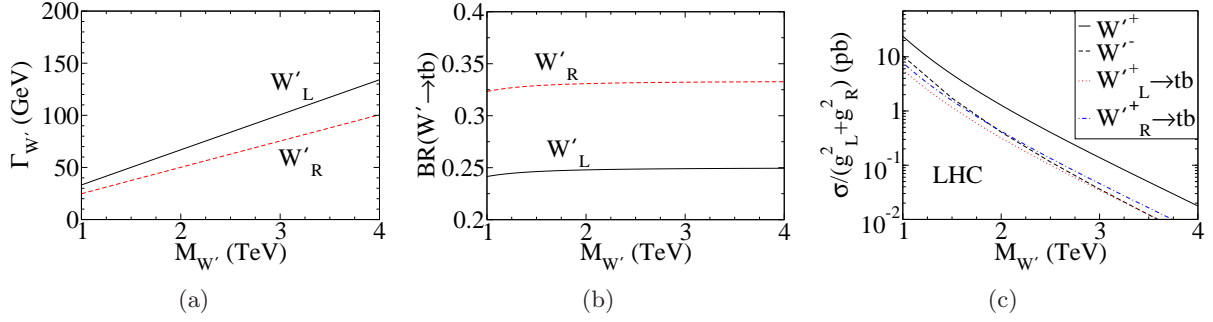

\centering
\subfigure[]{
	\includegraphics[width=0.3025\textwidth]{wpwidth.eps}%,height=0.2\textheight]{wpwidth.eps}
	\label{wpwidth.FIG}
}
\subfigure[]{
	\includegraphics[width=0.3025\textwidth]{BR.eps}%,height=0.2\textheight]{BR.eps}
	\label{BR.FIG}
}
\subfigure[]{
	\label{mwprod.FIG}
	\includegraphics[width=0.31\textwidth]{mwprod.eps}%,height=0.2\textheight]{mwprod.eps}
	}
\caption{ (a) The total decay width, (b) the branching fraction of $\wpri\rightarrow t\bar{b}$, 
and (c) the cross sections of the production of ${\wpri}^\pm$  at the 14 TeV LHC with and without its 
subsequent decay to $t\bar{b}$,
as a function of $\wpri$ mass for pure gauge eigenstates $W'_L$ and $W'_R$. }
\label{proddec.FIG}
\end{figure}

The partial width for $\wpri$ decaying to a pair of quarks is
\bea
\nonumber
\Gamma(W'\rightarrow \bar{q}q') &=& 3{g_2}^2({g^{qq'}_L}^2+{g^{qq'}_R}^2)\frac{\mwpri}{48\pi},
\qquad (m_q=m_{\bar q}=0) \\
\Gamma(W'\rightarrow tb) &=& 3{g_2}^2({g^{tb}_L}^2+{g^{tb}_R}^2)\frac{\mwpri}{48\pi}\bigg{(}1-\frac{m^2_t}{\mwpri^2}\bigg{)}\bigg{(}1-\frac{m^2_t}{2 \mwpri^2}-\frac{m^4_t}{2 \mwpri^4}\bigg{)}, \qquad (m_{b}=0).
\label{parttop.EQ}
\eea
Here and henceforth, we generically denote the $W'$ with  both left- and  right-handed couplings
$g_L$ and $g_{R}$, and set the gauge coupling strength to that of the SM SU(2), $g_2$. 
The partial widths of the $\wpri$ to quarks is symmetric under $R\leftrightarrow L$. 
However, its decay to the leptons will depend upon the lepton spectrum and flavor mixing
for a given model. If assuming the SM particle content and no heavy leptons $(N,\ L)$ below the threshold
$M_{W'} < M_L+M_{N}$, then one only has 
\beq
\Gamma(W'_{L}\rightarrow \ell \nu_i)={g_2}^2{g^{l\nu}_L}^2\frac{\mwpri}{48\pi},
%\Gamma(W'_{R}\rightarrow \ell N_i)={g_2}^2{g^{l N}_L}^2\frac{\mwpri}{48\pi}.
\eeq
where $\nu_{i}$ are the three SM-like light neutrinos.
Therefore, the pure gauge eigenstates $W'_L$ and $W'_R$ have different decay widths, given by
\bea
\Gamma_{W'_R}&=&\frac{{g_2}^2 g_R^2 \mwpri}{16\pi}\bigg{[}2+\bigg{(}1-\frac{m^2_t}{\mwpri^2}\bigg{)}\bigg{(}1-\frac{m^2_t}{2\mwpri^2}-\frac{m^4_t}{2\mwpri^4}\bigg{)}\bigg{]}\\
\Gamma_{W'_L}&=&\frac{{g_2}^2 g_L^2 \mwpri}{16\pi}\bigg{[}3+\bigg{(}1-\frac{m^2_t}{\mwpri^2}\bigg{)}\bigg{(}1-\frac{m^2_t}{2\mwpri^2}-\frac{m^4_t}{2\mwpri^4}\bigg{)}\bigg{]}.
\eea
Figure~\ref{wpwidth.FIG} shows the total width of the pure gauge eigenstates $W'_{L,R}$ as a function of $\mwpri$.
The $W'_L$ width is consistently larger than $W'_R$ since it has an additional decay channel to SM leptons,
although the partial width for $\wpri\rightarrow t\bar{b}$ is the same for both the left-handed and right-handed cases.
Figure \ref{BR.FIG} shows the branching fraction  for $\wpri\rightarrow t\bar{b}$ as a function of $\mwpri$,
%\beq
BR$(W'\rightarrow t\bar{b})= {\Gamma(W'\rightarrow t\bar{b})}/{\Gamma_{W'}}. $
%\eeq
It is smaller for the left-handed $\wpri$.  

To calculate the production cross section for 
the pure gauge eigenstates $\wpri\rightarrow t\bar{b}$ we used the narrow width approximation:
\beq
\sigma(pp\rightarrow W'\rightarrow t\bar{b})\approx\sigma(pp\rightarrow W')\rm{BR}(W'\rightarrow t\bar{b}),
\label{eq:narrow}
\eeq
where $\sigma(pp\rightarrow W')$ is the on-shell production cross section of the $W'$.
Figure~\ref{mwprod.FIG} shows the production cross section for the pure gauge eigenstate $W'^+$ (solid line) and ${\wpri}^-$ (dashed line) before the $\wpri$ decays, 
and the cross section for $pp\rightarrow {\wpri}^+_L\rightarrow t\bar{b}$ (dotted line) and $pp\rightarrow {\wpri}^+_R\rightarrow t\bar{b}$ (dash-dotted line) at the 14 TeV LHC, where the cross section is calculated according to Eq.~(\ref{eq:narrow}).  Since the valence quarks of the proton are two $u$ quarks and one $d$ quark, a process with an initial-state containing $u$ quarks will have a larger production cross section when compared to a similar process with an initial-state containing $d$ quarks.  Hence, the ${\wpri}^+$ production cross section is greater than the ${\wpri}^-$ production cross section.  To be precise, the cross section for ${\wpri}^+$ production ranges from around $2.3$ times to $4.2$ times larger than the ${\wpri}^-$ cross section for a $\wpri$ mass between $1$ and $4$ TeV.  
Also, the cross section for $pp\rightarrow {\wpri}^+_L\rightarrow t\bar{b}$ is smaller than that of 
$pp\rightarrow {\wpri}^+_R\rightarrow t\bar{b}$ due to the differences in the branching fractions. 

The presence of $W'_L \leftrightarrow W'_R$ mixing would alter the above
characteristics since the observed mass eigenstate would be a linear combination of 
the gauge eigenstates $W'_L$ and $W'_R$. Thus, determining the chiral couplings of the 
$W'$ state would help in unravelling the presence of such mixings.

%%%%%%%%%%%%%%%%%%%%%%%%
\subsection{Event Characteristics and Reconstruction}
To determine the $\wpri$ chiral coupling from angular distributions, one would  like to know the charge of the
final state particle  produced. We thus focus on the leptonic mode of the top-quark decay.
The neutrino is unobserved but kinematical constraints can be used to infer its 4-momentum.
The transverse momentum of the neutrino can be found by momentum conservation from the observable
particles:
\begin{eqnarray}
\mathbf{k}_{\nu T}=-(\mathbf{k}_{\ell T}+\mathbf{k}_{3T}+\mathbf{k}_{ 2 T})
\end{eqnarray}
where the momentums are as labeled in Fig.~\ref{qq2Wp2tb.FIG},  and the subscript $T$ labels the transverse momentum.  
However, the neutrino's longitudinal momentum cannot be determined through momentum conservation due to 
the unknown boost of the partonic c.m.~frame. 
It can be inferred through the on-shell condition for the $W$-mass
\beq
M^2_W=(k_\nu+k_{\ell })^2 .
\label{eq:wmass}
\eeq
Solving this quadratic equation for the neutrino longitudinal momentum 
leads to a two-fold ambiguity.  We further impose the on-shell condition for the top-quark
\begin{eqnarray}
m^2_t=(k_\nu+k_\ell +k_3)^2 .
\end{eqnarray}
In principle, we do not know which of the two $b$-quarks is coming from the top-quark decay.  
The previous two-fold ambiguity is now a four-fold ambiguity. 
For each of these four possibilities we evaluate the top mass 
and pick the solution that is closest to the measured top mass.
This resolves the four-fold ambiguity, determines the longitudinal component of the neutrino, and  identifies which $b$-quark is the one resulting from top decay.
The event can be completely resolved and the $t\bar b$ c.m.~frame is reconstructed. 
We have checked against our theoretical calculation that in this method the reconstruction is correct 
about $99.99\%$ in identifying the $b$-quark, and in identifying the correct neutrino momentum.

Another technique one could use is to assume that the $b$-jet closer (in opening angle) 
to the lepton is the one from top decay. This is based on the fact that the top-quark from a heavy $W'$
decay is rather energetic, so that its decay products are collimated. 
This method of reconstruction correctly identifies $b$-quark $93.5\%$ of the time 
and the neutrino momentum 96.7\% of the time. We will adopt the first method for the event reconstruction.

\begin{figure}[tb]
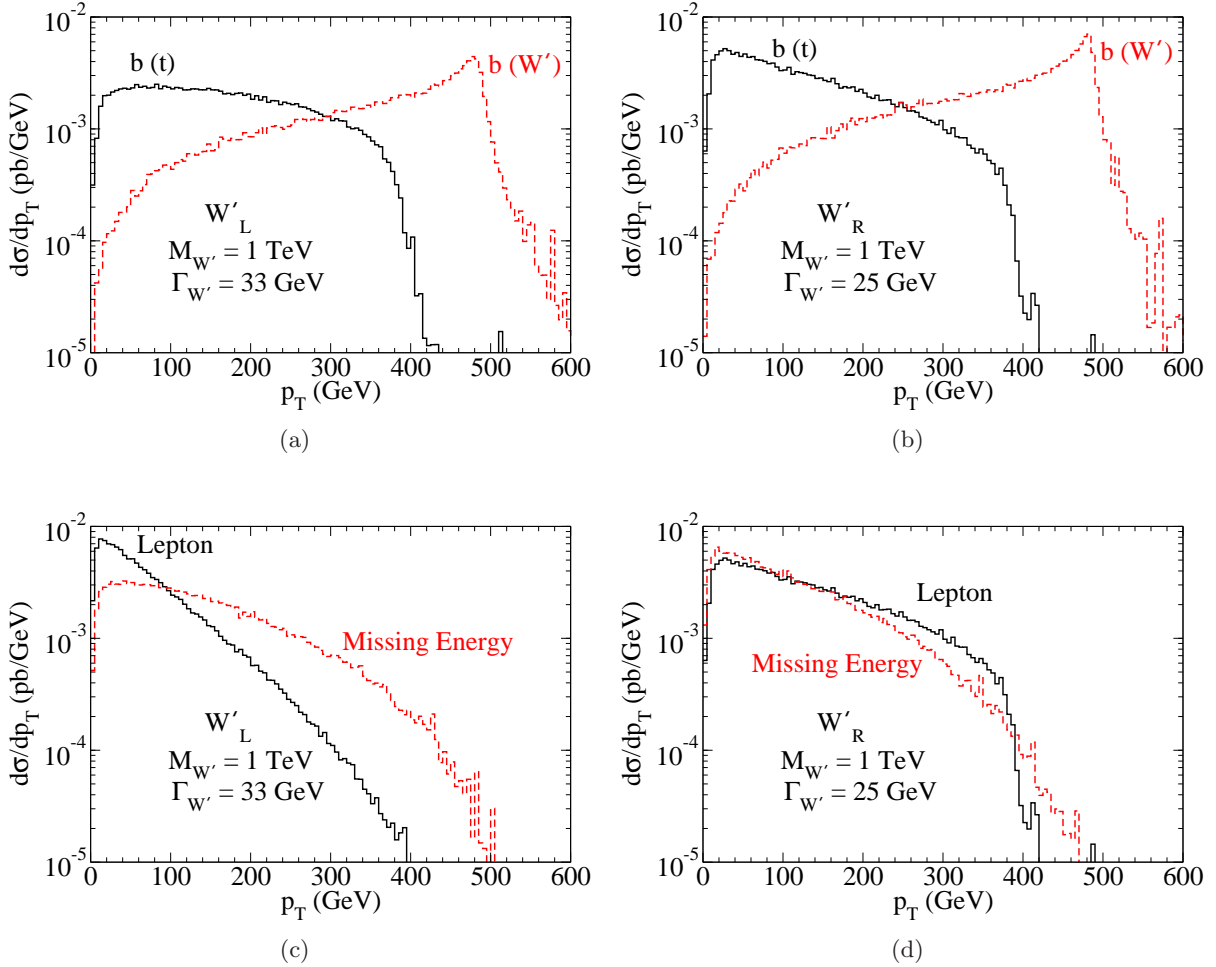

\centering
\subfigure[]{
	\label{ptdistLHB.FIG}
	\includegraphics[width=0.47\textwidth]{ptbotLHNC.eps}
	}
\subfigure[]{
	\includegraphics[width=0.47\textwidth]{ptbotRHNC.eps}
	\label{ptdistRHB.FIG}
}\vspace{.15in}\\
\subfigure[]{
	\label{ptdistLHNC.FIG}
	\includegraphics[width=0.47\textwidth]{ptlepLHNC.eps}
	}
\subfigure[]{
	\includegraphics[width=0.47\textwidth]{ptlepRHNC.eps}
	\label{ptdistRHNC.FIG}
	}
\caption{Transverse momentum distributions 
in $pp\to W'\to t\bar{b}\to b\bar{b}\ \ell^+\nu_\ell$ production at the LHC  for $M_{W'}=1$ TeV,
for the two $b$-jets from (a) the ${\wpri}_L^+$ and (b) ${\wpri}_R^+$; 
and for the lepton and missing energy from (c) the ${\wpri_L}^+$ and (d) ${\wpri_R}^{+}$.
No acceptance cuts are imposed and the  SM $W$ contribution is not included.}
\label{ptdist.FIG}
\end{figure}

\begin{figure}[tb]
\begin{center}
\subfigure[]{
\includegraphics[width=0.29\textwidth,clip=true]{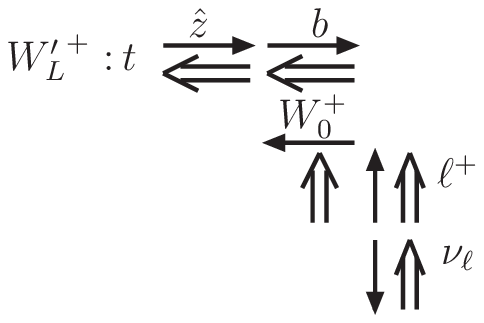}
%\label{LspinRH.FIG}
        }
\subfigure[]{
\includegraphics[width=0.29\textwidth,clip=true]{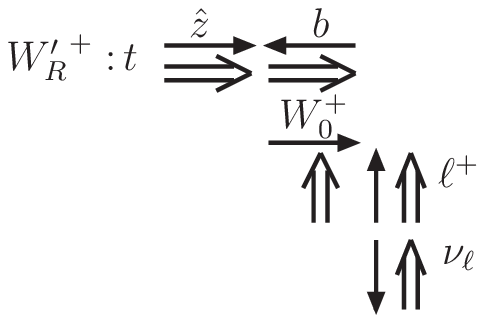}
%\label{LspinRH.FIG}
        }\\
\subfigure[]{
\includegraphics[width=0.29\textwidth,clip=true]{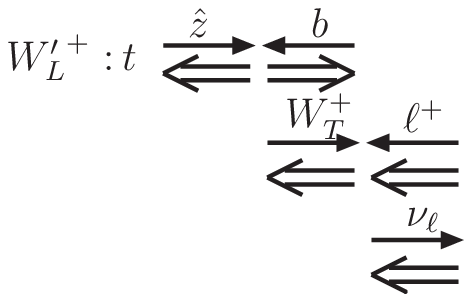}
%\label{LspinRH.FIG}
        }
\subfigure[]{
\includegraphics[width=0.29\textwidth,clip=true]{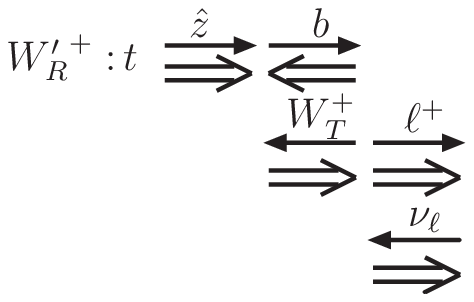}
\label{LspinLH.FIG}
        }
\caption{Helicity and spin correlations in the chains $t_{L,R} \to b W^{+} \to b\ \ell^{+}\nu_{\ell}$ 
from $\wpri_L$ decay in  (a),~(c); and from $\wpri_R$ decay in (b),~(d).
Figures (a) and (b) are for longitudinally polarized SM $W$'s, and Figs.~(c) and (d) are for 
transversely polarized SM $W$'s.  The decay goes from left to right as labeled by the particle names. 
The momenta (single arrow lines) and spins (double arrow lines) are in the parent rest frame 
in the direction of the top-quark's motion ($\hat{z}$) in the $\wpri$ rest frame.
}
\label{Lspin.FIG}
\end{center}
\end{figure}

Once we have determined the 4-momentum of the neutrino and which $b$-quark results from the top decay, we have  
fully reconstructed the system and can proceed to form various kinematical quantities
which have sensitivity to the $\wpri$ chiral couplings. 
In the following calculations, spin correlations in the decay chains have been fully implemented. 
Let us first examine the transverse momentum distribution for various final state particles. 
We recall that for a parent particle
of mass $M$ decaying to two light particles, the transverse momentum of the final state particle develops a Jacobian peak near
$M/2$. This peak will be subsequently smeared by the transverse motion of the parent particle. 
For the sake of signal illustration in this section, 
no acceptance cuts are imposed and  the  SM contribution via $W$ exchange is not included here.
In Figs.~\ref{ptdist.FIG}~(a) and \ref{ptdist.FIG}~(b), 
we show the transverse momentum of the two $b$-jets for the $W'_L$
and $W'_R$, respectively. The harder $b$ jet clearly shows the above kinematical feature from a heavy $W'$ decay.
It is interesting to notice the slight difference for the softer $b$ jet in these two panels: $W'_L$ leads to a harder $b$ jet
than $W'_R$. This is a natural consequence of the top-quark spin correlation as illustrated in Fig.~\ref{Lspin.FIG}, where the 
decay chains of a polarized top-quark of L,R helicities are depicted. 
Figures \ref{Lspin.FIG}~(a) and \ref{Lspin.FIG}~(b) illustrate the dominant decay chain with a longitudinally polarized 
$W$ boson ($W^{+}_{0}$),
whereas the accompanying $b$ quark is boosted along (against) its motion by a parent top-quark 
in the case of ${\wpri_L}^{+}$ (${\wpri_R}^{+}$).
The transverse momentum distributions of the lepton and missing neutrinos 
for the left-handed [Fig.~\ref{ptdistLHNC.FIG}] and right-handed [Fig.~\ref{ptdistRHNC.FIG}] $\wpri$
can be understood similarly 
as seen in Fig.~\ref{Lspin.FIG}.
The charged leptons tend to be softer in the left-handed case than those in the right-handed case, as seen from the
correlation in Figs.~\ref{Lspin.FIG}~(a) and \ref{Lspin.FIG}~(b). Although subleading with about $30\%$ contribution from the
transversely polarized $W$ as shown by Figs.~\ref{Lspin.FIG}~(c) and \ref{Lspin.FIG}~(d),
the successive boosts of the top and the $W$ in the same (opposite) direction for the $\wpri_L$ ($\wpri_R$)
%those diagrams with double parent-boosts 
lead to a softer (harder) charged lepton distribution and a
harder (softer) $\etmiss$ distribution as seen in 
Figs.~\ref{ptdistLHNC.FIG} and \ref{ptdistRHNC.FIG}.
Finally, for the case of ${\wpri_R}^{+}$, 
the charged lepton transverse momentum distribution is only slightly harder than that of missing neutrinos, which
indicates the slightly larger boost of the top-quark in one direction compared to the 
boost of the $W$ in the opposite direction.

%%%%%%%%%%%%%%%%%%%%%%%%%%%%%%%%%%%%%
\subsection{Signal Selection and Background Suppression}

We perform an analysis including SM contribution  
\begin{equation}
pp \to W^{+*} \to t \bar{b} \to \bar{b}\ b\ell^{+} \nu ,
\label{eq:signal}
\end{equation}
which serves as the irreducible SM background. Interference between the $\wpri$ and SM diagrams is
fully implemented. 

To be as realistic as possible, we simulate detector effects by smearing the lepton 
and jet energies according to the assumed Gaussian resolution parametrization
\bea
\frac{\sigma(E)}{E}=\frac{a}{\sqrt{E}}\oplus b,
\label{eq:smear}
\eea
where $\sigma(E)/E$ is the energy resolution, $a$ is a sampling term, $b$ is a constant term, $\oplus$ represents addition in quadrature, and all energies are measured in GeV.  For leptons we take $a=5\%$ and $b=0.55\%$, and for jets we take $a=100\%$ 
and $b=5\%$ \cite{Ball:2007zza}.  After smearing we apply the basic acceptance cuts:
\bea
p_T( \ell)\, >\, 20 \, {\rm GeV}, ~\eta(\ell)\, < \, 2.5, \qquad
p_T( j )\, >\, 50 \, {\rm GeV}, ~ \eta( j )\, < \, 3.0,\qquad
\etmiss \, > \, 25 \, {\rm GeV}.
\label{cuts1} 
\eea
The angular separation between particle $i$ and particle $j$ is defined by
\beq
\Delta R_{ij}=\sqrt{\Delta\phi^2_{ij}+\Delta\eta_{ij}^2},
\eeq
where $\Delta\phi_{ij}$ is the difference between the particles' azimuthal angles, and $\Delta\eta_{ij}$ is the difference between the particles' rapidities.  
The top-quark and the $\bar{b}$ resulting from the $\wpri$ will be back-to-back in the transverse plane.  Hence, the angular separation between the $b$ resulting from the top-quark decay and the $\bar{b}$ resulting from the $\wpri$, $\Delta R_{bb}$, is peaked near $\pi$.  On the other hand, since the top is highly boosted and its decay products collimated, the angular separation between the lepton and $b$ from the top decay, $\Delta R_{\ell b}$, is peaked at a low value.   Hence, a stringent cut on $\Delta R_{\ell b}$ would cut out most of our signal.  We thus impose the cuts
\bea
\Delta R_{\ell b}>0.3,~~~~\Delta R_{bb}>0.4\,
\label{cuts3}
\eea
where $\Delta R_{\ell b}$ is the $\Delta R$ for the lepton and the $b$-jet resulting from the top decay, and $\Delta R_{bb}$ is the $\Delta R$ for the two $b$-jets.

\begin{figure}[tb]
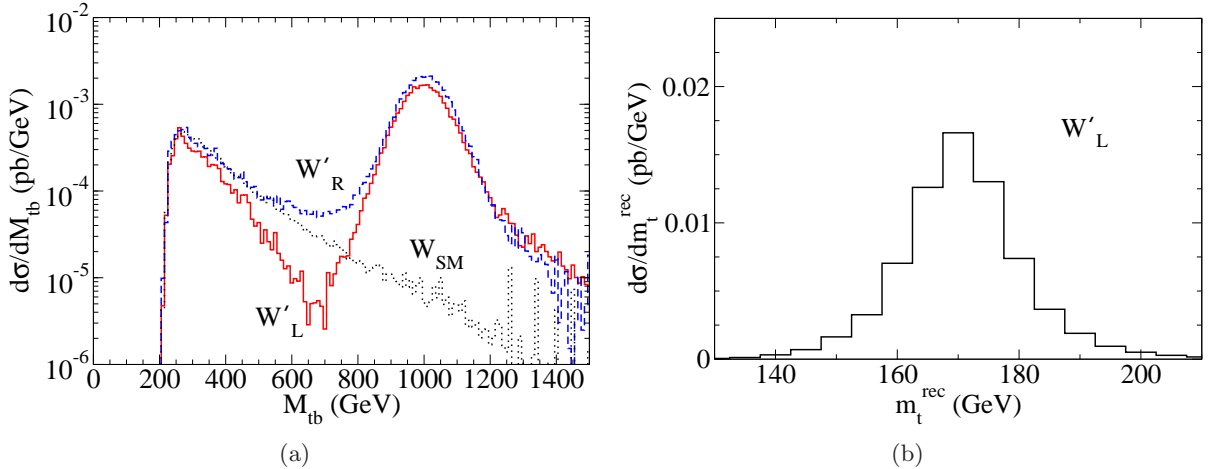

\centering
\subfigure[]{
	\label{TotInvM.FIG}
	\includegraphics[width=0.47\textwidth]{LHC14invmass.eps}
	}
\subfigure[]{
	\includegraphics[width=0.47\textwidth]{mtoprec.eps}
	\label{mtoprec.FIG}
}
\caption{(a) The invariant-mass distribution of $tb$ system for 
$pp\rightarrow W/W'\rightarrow t\overline{b}\rightarrow b\overline{b}\ell^+\nu$ production at the LHC with $M_{W'}=1$ TeV
with contributions from $W+W'_L$ (solid line), $W+W'_R$ (dashed line), and the standard model $W$ (dotted line); 
and (b) the invariant-mass distribution for the reconstructed top-quark. }
\end{figure}

%As the top transverse momentum increases it becomes harder to resolve the the lepton and $b$-jet until above a transverse momentum of about $1350$~GeV it becomes more likely that they are not resolved.  
%

Once we have fully reconstructed the event 
after momentum smearing and including the cuts in Eqs.~(\ref{cuts1}) and (\ref{cuts3}),  
we can further reconstruct the partonic c.m.~frame, invariant-mass $M_{t\bar{b}}$, 
and the top-quark mass $m^{\rm{rec}}_t$. 
The most convincing signal identification would be the reconstruction of the mass peak. 
Figure~\ref{TotInvM.FIG} shows the system invariant-mass distributions for a $1$ TeV $W'_L$ and $W'_R$, including the SM contribution.  
As can be seen the $W'_L$ and SM $W$ destructively interfere before the resonance peak, while the $W'_R$ and SM $W$ constructively interfere before the peak. 
 Also, near the resonance the $W'$ signal dominates over the SM $W$.  At the $\wpri$ peak, the small difference between the $\wpri_R$ and $\wpri_L$ invariant-mass distributions is due to their different decay widths, resulting in slightly different decay branching fractions to the
 final states.
%, although it is not possible to measure those couplings from the invariant-mass distribution.  

As pointed out previously in Ref.~\cite{Rizzo:2007xs}, since the $\wpri_L$ ($\wpri_R$) destructively (constructively) interferes with the SM $W$ before the resonance peak, 
it may be possible to determine if the $\wpri$ has left-handed or right-handed couplings by observing the interference region in the total partonic c.m.~frame mass distribution.  Our proposed observables are defined around the $\wpri$ resonance peak.  Since the interference region has a much lower cross section than the resonance peak, our proposed observables will have a better ability to determine the $\wpri$ chiral couplings once a resonance is observed.
 
Figure~\ref{mtoprec.FIG} shows the reconstructed top mass distribution from  a $1$ TeV $W'_L$ plus the SM $W$ after momentum smearing, the cuts in Eq.~(\ref{cuts1}), and event reconstruction.  The reconstructed top mass distribution stays highly peaked near the measured top mass. The apparent width is dominated by the jet energy resolution. The top-quark reconstructed mass distribution for the $\wpri_R$ is largely the same. 

Motivated by the distributions in Figs.~\ref{TotInvM.FIG} and~\ref{mtoprec.FIG}, we apply cuts on the reconstructed system invariant-mass, $M_{t\bar{b}}$, and the reconstructed top-quark mass, $m^{\rm{rec}}_t$:
\bea
\mwpri\, -\, 100\, {\rm GeV}\, \leq  \, &M_{t\bar{b}}& \leq \mwpri\, +\, 100\, {\rm GeV}, \nonumber \\
%1900\, {\rm GeV} \, \leq  \, M_{t\bar{b}} \leq 2100\, {\rm GeV} \qquad ({\rm for}\ \mwpri = 2~{\rm TeV}),
m_t\, -\, 20\, {\rm GeV} \, \leq  \, &m^{\rm{rec}}_t& \leq m_t\, +\, 20\, {\rm GeV} \ .
\label{cuts2}
\eea

\begin{table}[tb]
\caption{Signal cross section $pp\to t\bar{b}\to b\bar{b}\ \ell^+\nu_\ell$ where $\ell=e^+$ 
or $\mu^+$ for $\mwpri = 1$~TeV with and without the SM $W$ contribution, 
with consecutive cuts as listed.
\label{csSBirr1TeV.TAB}}
\begin{center}
\begin{tabular}{|l|c|c|c|c|}  \hline
$\sigma$(pb)                                  & $W+W'_L$  & $W+W'_R$ & $W'_L$ & $W'_R$ \\ \hline\hline
No cuts or smearing                             &  1.1         &  1.4      &0.67    &0.90  \\ \hline
No Cuts                                       &  0.92         &  1.2      &0.57    &0.75   \\ \hline
Cuts Eq. (\ref{cuts1})                       &0.38           &  0.51      &0.32    &0.42   \\ \hline
~~+Eq.~(\ref{cuts3})                            &  0.35         &0.46        &0.30    &0.37   \\ \hline
~~+Eq.~(\ref{cuts2})                            &0.22         &0.29       &0.22    &0.29    \\ \hline
~~+Eq.~(\ref{cuts4}) \& tagging 1 $b$-jet      &0.11             & 0.16          &0.11   &0.15    \\ \hline \hline
~~+tagging 2 $b$-jets without Eq.~(\ref{cuts4})              &0.070        &0.10       &0.070   &0.10    \\ \hline
\end{tabular}
\end{center}
\end{table}

In Table~\ref{csSBirr1TeV.TAB} 
 we show the effects of the cuts on the signal with and without  the SM $W$ contribution for a $1$ TeV $\wpri$. As can be seen, as the cuts progress up to Eq.~(\ref{cuts2}),  the SM $s$-channel $W$ background disappears and we are left with our resonance signal.

Besides the SM $s$-channel $W$ background, there is also the irreducible QCD background of 
$b\bar{b}W^+$.  
A QCD jet from a light quark or a gluon may be misidentified as a $b$ quark and
hence results in reducible backgrounds $jj W^+$. 
Furthermore, another large background comes from the $t$-channel single top-quark production, 
$W^{*} b \to t$, which can be considered at a more fundamental level as
\begin{equation}
q\ g \to q'\ t\ \bar b .
\label{eq:back}
\end{equation}
In fact, both $q'$ and $\bar b$ in the final state move in the forward direction with a low transverse momentum
(at the order of $M_{W}/2$ and $m_{b}$, respectively).
If we envision a search for exclusive signal $t\bar b$ as Eq.~(\ref{eq:signal}), the above backgrounds 
constitute two classes of final states denoted by
%\begin{equation}
$W^+ g \to t\bar b$ and $b q \to t q'$,
%\end{equation}
where $q'$ could fake $\bar b$.
Hence, in the following analysis we separately consider those two final states and apply no cuts on the
additional jet (nor is it used in the reconstruction). Our background estimate is on the conservative side
since we could apply an explicit jet-vetoing cut on the extra forward jet in Eq.~(\ref{eq:back}). 

The background events are first smeared according to Eq.~(\ref{eq:smear}) and then run through the reconstruction algorithm.  Although the lepton and neutrino originate from $W$ decay, for some events solving Eq.~(\ref{eq:wmass}) for the neutrino longitudinal momentum does not result in physical solutions.  Analyzing Eq.~(\ref{eq:wmass}), one can determine that for a solution to exist the following condition must be met:
\begin{eqnarray}
M^2_W \ge  2\ k_{\nu T}\ k_{\ell T}\ \left(1-\cos(\phi_\ell-\phi_\nu)\right),
\label{reccnd.EQ}
\end{eqnarray}
where $\phi_i$ is the azimuthal angle of particle $i$.  If the missing energy and lepton transverse momenta are near $M_W/2$ and nearly back-to-back, once energy smearing is applied it is possible that $4k_{\nu T}k_{\ell T}\ge M^2_W$.  Hence, for these events Eq.~(\ref{reccnd.EQ}) is not satisfied and there are no physical solutions for the neutrino longitudinal momentum.  Events for which solutions do not exist are rejected.

Our background cross sections with consecutive cuts are listed in Table~\ref{csBred.TAB}.  As can be seen, after reconstruction and the cuts in Eqs.~(\ref{cuts1}) and (\ref{cuts3}), the backgrounds still dominate over our signal.  Applying the invariant-mass cuts in Eq.~(\ref{cuts2}) suppresses  the $jjW,\ tj,\ W g$ 
backgrounds by one to two orders of magnitude and the QCD $b\bar{b}W$ background by three orders of magnitude while only affecting our signal by $20-30\%$. Despite the highly effective kinematical cuts, the 
$jjW$ background cross section is still more than three times as large as that of the signal.  

To further suppress this background, we consider the $b$-tagging.
The probability that a $b$-jet is correctly identified is $60\%$ while the faked rates of mistagging a light jet as a $b$-jet is transverse momentum dependent and we adopt them from \cite{Ball:2007zza}:
\begin{eqnarray}
\varepsilon_\ell=\left\{
\begin{array}{l l}
{1\over 150}  &\quad p_T<100~\rm{GeV}\\
\frac{1}{450}\bigg{(}\frac{p_T}{25~\rm{GeV}}-1\bigg{)}&\quad 100~{\rm GeV}<p_T<250~\rm{GeV} \\
{1\over 50}  &\quad~p_T>250~\rm{GeV}\\
\end{array}\right .
\end{eqnarray}

\begin{table}[tb]
\caption{SM background cross sections with consecutive cuts optimized  to a $1$ TeV $\wpri$ signal. 
\label{csBred.TAB} }
\begin{center}
\begin{tabular}{|l|c|c|c|c|}\hline
$\sigma$ (pb)               &$jjW^+$      & $bq\to tj$   & $W^+g \rightarrow t\bar{b}$   &$b\bar{b}W^+$      \\ \hline \hline
%Cuts Eqs.~(\ref{cuts1},\ref{cuts3}) &$1.0$ &$94$ &$0.22$ \\ \hline
Cuts Eqs.~(\ref{cuts1},\ref{cuts3})            &$96$               &  $2.3$                       &$0.56$                        & $0.18$             \\\hline
~~+Eq.~(\ref{cuts2})                           &$1.3$              & $0.080$                      &$4.4\times10^{-3}$            &$4.6\times10^{-4}$ \\\hline
~~+Eq.~(\ref{cuts4}) \& tagging 1 $b$-jet      &$2.5\times10^{-3}$ & $6.9\times10^{-3}$           &$1.5\times10^{-3}$            &$2.7\times10^{-4}$ \\ \hline \hline
~~+tagging 2 $b$-jets without Eq.~(\ref{cuts4})&$1.4\times10^{-4}$ & $2.5\times10^{-4}$            &$1.6\times10^{-3}$            &$1.6\times10^{-4}$ \\\hline
\end{tabular}
\end{center}

\end{table}

Finally we note, as seen from Figs.~\ref{ptdistLHB.FIG} and \ref{ptdistRHB.FIG}, that 
the $b$-jet originating from the $\wpri$ will have a large transverse momentum peaking around $M_{W'}/2$.  
This highly energetic jet would make the $b$-tagging less efficient and result in a larger fake rate as well. 
Instead, we could consider to trade this second $b$-tagging for the harder jet with a higher transverse momentum cut
\bea
p_{T,\rm{max}}(j)~>~300~\rm{GeV}
\label{cuts4}
\eea
The second to last row of Table \ref{csSBirr1TeV.TAB} (\ref{csBred.TAB}) shows the effect of the single $b$-tagging plus the cut in Eq.~(\ref{cuts4}) on the harder jet for the signal (background). 
As can be seen, tagging one $b$-jet and applying an additional $p_T$ cut, Eq.~(\ref{cuts4}), further suppresses the $W$ plus 
two-jet background by three orders of magnitude and the $t$-channel top-quark background by one order of magnitude and only decreases the signal rate by half.  
For the sake of comparison, the last row of Table~\ref{csSBirr1TeV.TAB} (\ref{csBred.TAB}) provides the cross sections obtained from tagging both $b$-jets using the efficiencies above and without the cut in Eq.~(\ref{cuts4}).
Tagging both jets as $b$-jets suppresses the $Wjj$ by four orders of magnitude and $t$-channel top-quark backgrounds by two orders of magnitude, while the signal rate is decreased, depending on the $b$-tagging efficiency for the high-energy $b$, by about two thirds.
Once the $b$-tagging is implemented, our signal for a 1 TeV $W'$ 
is larger than the background by at least an order of magnitude.

%%%%%%%%%%%%%%%%%%%%%%%%%%%%
\subsection{Heavier $\wpri$ and Highly Boosted Top Quark}

\begin{figure}[tb]
\centering
\subfigure[]{
        \includegraphics[width=0.30\textwidth]{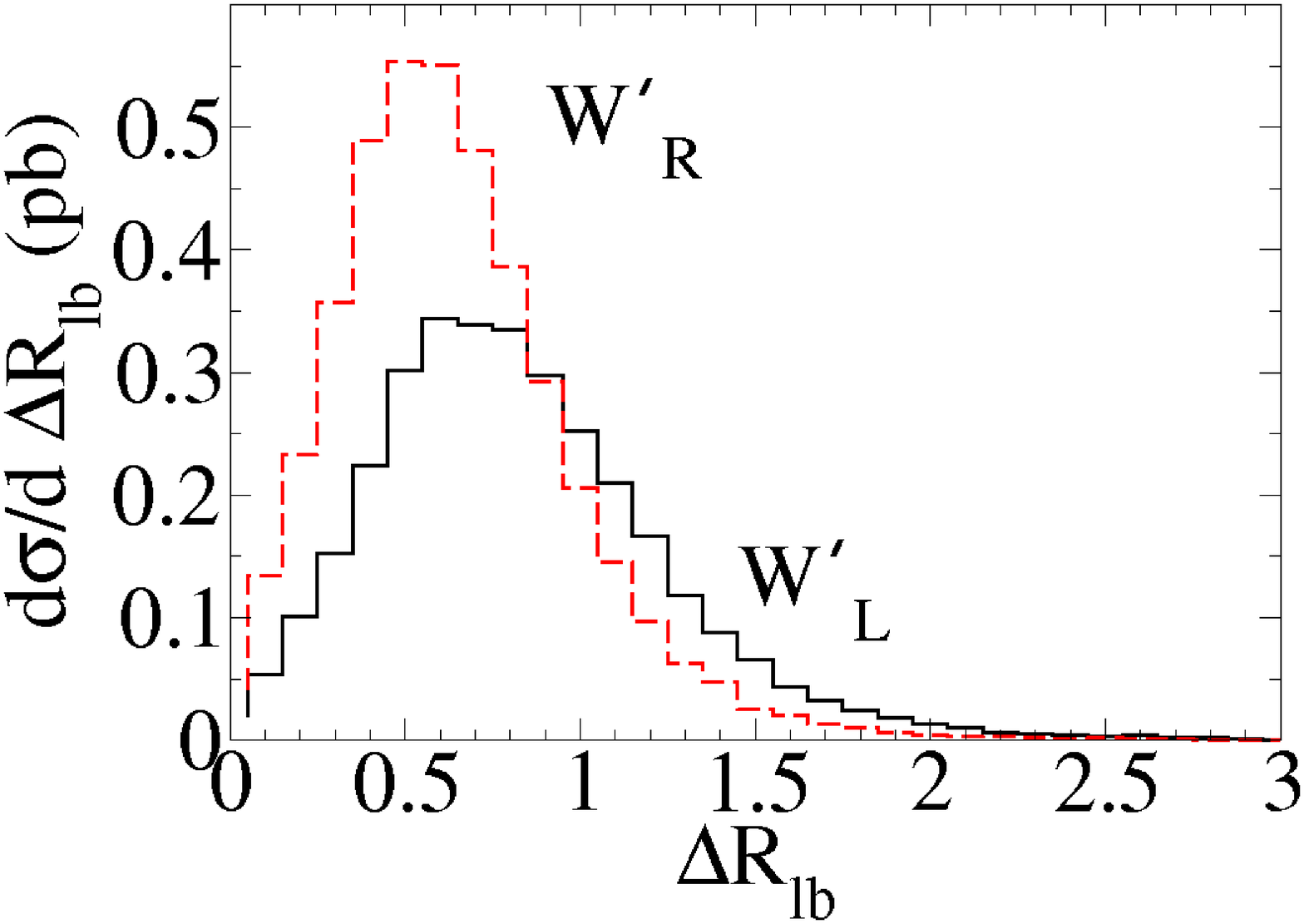}
        \label{rlb.FIG}
}
\subfigure[]{
        \includegraphics[width=0.31\textwidth]{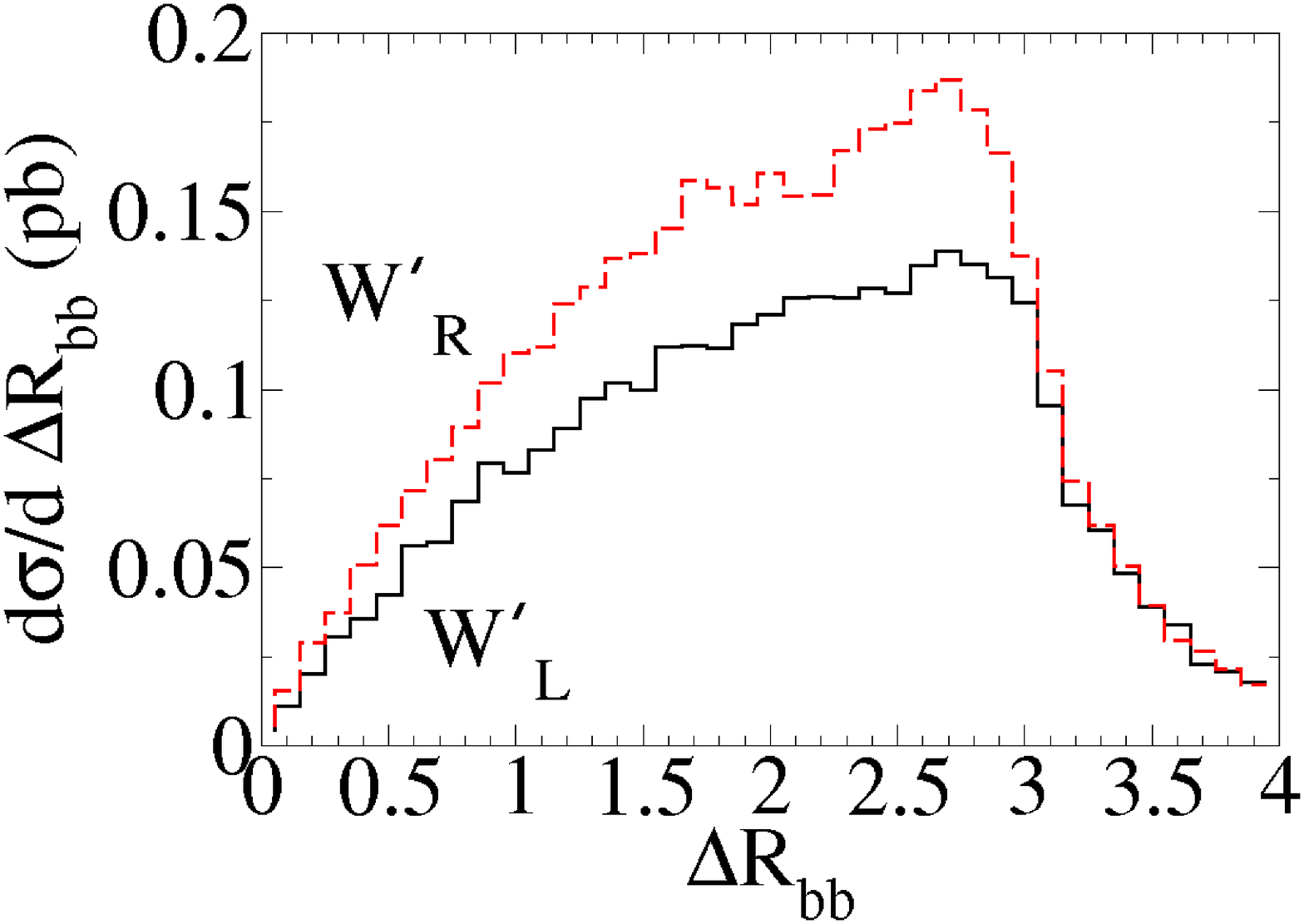}
        \label{rbb.FIG}
}
\subfigure[]{
        \includegraphics[width=0.3025\textwidth]{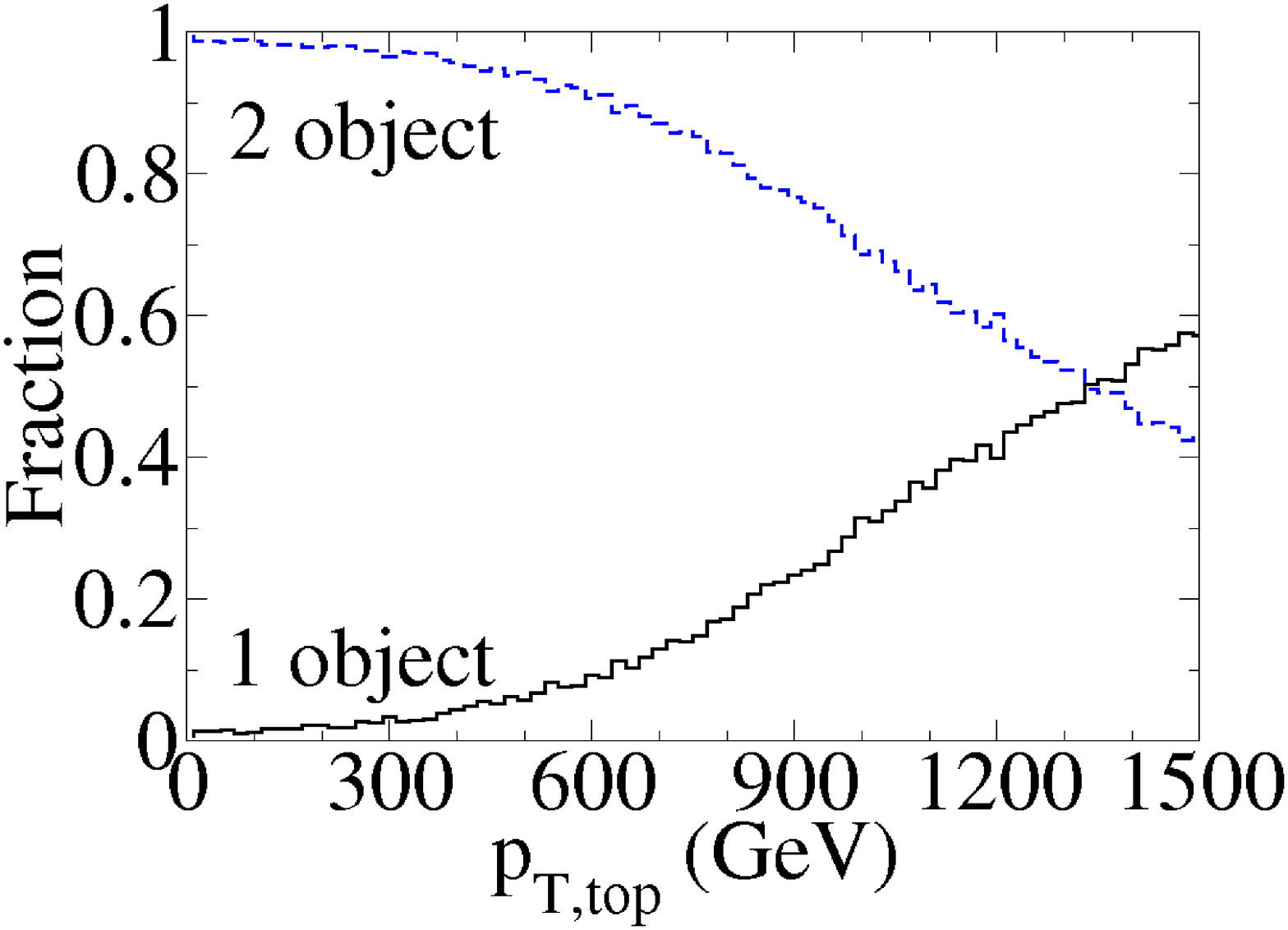}
        \label{perc.FIG}
}
\caption{$pp\to W'\to t\bar{b}\to b\bar{b}\ \ell^+\nu_\ell$ production at the LHC for $M_{W'}=1$ TeV
(a) for the distribution $\Delta R_{\ell b}$ between the lepton and the jet identified with the $b$, 
(b) for the distribution $\Delta R_{bb}$ between the two $b$-jets, 
and (c) for the 
fraction of events with the number of objects in the decay of the top-quark
versus the top transverse momentum.  }
\label{deltar.FIG}
\end{figure}

The energy of a top-quark from the $W'$ decay is roughly $\mwpri/2$. 
For a more massive $\wpri$, the top-quark  becomes very energetic and its decay products are more collimated.
Figure~\ref{rlb.FIG} shows the $\Delta R$ distribution for the lepton and the $b$-jet coming from the top decay and 
Fig.~\ref{rbb.FIG} shows the $\Delta R$ distribution for the two $b$-jets.  Both figures show the distributions for the purely right-handed and the purely left-handed chiral couplings between the SM fermions and a $1$~TeV $\wpri$ after smearing and the cuts in Eq.~(\ref{cuts1}).  The lepton and $b$-jet are coming from a boosted top and so should not have much angular separation; hence, the distribution in Fig.~\ref{rlb.FIG} is peaked at low $\Delta R_{\ell b}$, approximately at $2m_{t}/E_{tT}\approx 0.5$ for a 1 TeV
$W'$. 
Since one $b$-jet is coming from the $\wpri$ decay and the other from the top-quark, the two $b$-jets are expected to land on opposite sides of the detector and the distribution in Fig.~\ref{rbb.FIG} is peaked at large $\Delta R_{bb}$ near $\pi$. 

Let us define an isolated object with a separation cut $\Delta R>0.3$. 
Figure~\ref{perc.FIG} shows the fraction of events for which the $b$-jet from the top and the lepton can and cannot be resolved with respect to the top transverse momentum.
For the sake of illustration, 
we have used a left-handed $\wpri$ mass of $3$~TeV, although this distribution is largely independent of the $\wpri$ mass.  
As the top-quark transverse momentum approaches 1.3 TeV, about one-half of the $b\ell$ events looks like a single 
object \cite{Barger:2006hm}, and thus our scheme for the c.~m.~frame reconstruction is inapplicable. 
Various methods have been proposed  to identify a highly boosted top-quark as  single 
``fat top jet'' \cite{Djouadi:2007eg,Kaplan:2008tt}. 
Jet substructures from massive particle decays may help in  top-quark identification \cite{Almeida:2008yp}.

%%%%%%%%%%%%%%%%%%%%%%%%%%%%%%%%%%%%%%%
\section{$W'$ CHIRAL COUPLINGS FROM ANGULAR CORRELATIONS AT THE LHC}
\label{WpriAC.SEC}
In this section we identify several kinematical quantities 
that depend directly on the chiral couplings of the $\wpri$ to SM particles,
and show that each has a different dependence on the 
various chiral couplings of the $W'$ and can be used to determine the couplings.

\subsection{$W'$ Chiral Coupling From Top Quark Angular Distribution}
\label{TopAngDist.SEC}

In the partonic c.m.~frame, the polar angular distribution of a final state fermion $f$ 
via a $W'$ production and decay is given by 
\\
\begin{eqnarray}
\frac{d\hat{\sigma}}{d\cos\theta_f}(q\bar q'\to W'\to f\bar f')&=&\frac{g_2^4}{256\pi}({g^{qq'}_R}^2+{g^{qq'}_L}^2)({g^{ff'}_R}^2+{g^{ff'}_L}^2)\frac{\hat{s}\lambda^{1/2}(1,x^2_f,x^2_{f'})}{(\hat{s}-M^2_{W'})^2+\Gamma^2_{W'}M^2_{W'}}\nonumber\\
&\times& \bigg{[}1-(x^2_f-x^2_{f'})^2-8x_fx_{f'}\frac{g^{ff'}_Rg^{ff'}_L}{{g^{ff'}_R}^2+{g^{ff'}_L}^2}+\lambda(1,x^2_f,x^2_{f'})\cos^2\theta_f\nonumber\\
&+& 2\lambda^{1/2}(1,x^2_f,x^2_{f'})\frac{{g^{qq'}_R}^2-{g^{qq'}_L}^2}{{g^{qq'}_R}^2+{g^{qq'}_L}^2}\frac{{g^{ff'}_R}^2-{g^{ff'}_L}^2}{{g^{ff'}_R}^2+{g^{ff'}_L}^2}\cos\theta_f\bigg{]}
\label{eqn:wct}
\end{eqnarray}
where $\hat{s}$ is the partonic c.~m.~energy squared, 
$\theta_f$ the angle between the final state fermion and the 
initial-state quark in the c.~m.~frame, 
$x_f=m_f/\sqrt{\hat{s}}$, and 
$\lambda(x,y,z)=(x-y-z)^2-4yz$ is the basic two-body kinematic function.

Since the LHC is a $pp$ collider, the quark can come equally from either proton, and a 
parity-violating angular distribution will be symmetrized, 
unless we distinguish the direction of the quark from that of the antiquark.
This can only be achieved approximately based on the argument 
%on a statistical basis by making use of the fact 
that an initial-state quark on average has a higher momentum fraction than an initial-state antiquark, 
since the former is a valence quark in the proton and the latter a sea-quark.  
Hence the quark and ${\wpri}^{\pm}$ are on average moving in the same direction.  

\begin{figure}[tb]
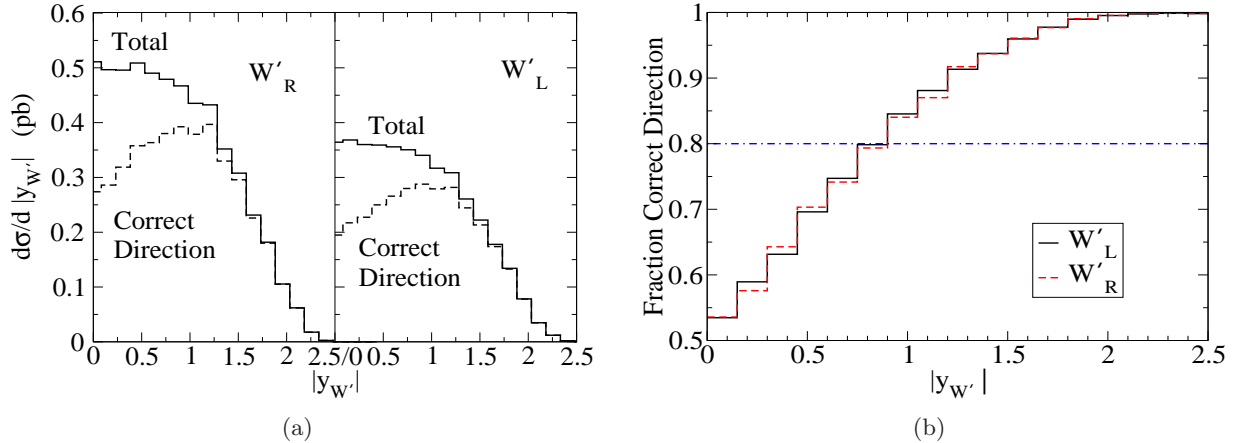

\centering
\subfigure[]{
	\label{yWp.FIG}
	\includegraphics[width=0.47\textwidth,height=0.231\textheight]{yWpNSmass.eps}
	}~~
\subfigure[]{
	\includegraphics[width=0.47\textwidth,height=0.231\textheight]{yWpfracNSmass.eps}
	\label{yWpfrac.FIG}
}
\caption{
$pp\to W'\to t\bar{b}\to b\bar{b}\ \ell^+\nu_\ell$ production at the LHC for $M_{W'}=1$ TeV versus
the reconstructed system rapidity $|y_{W'}|$,  (a) for all reconstructed events (solid), 
and events in which the reconstructed c.m.~frame and the quark are moving in the 
same direction (dashed, labeled as ``correct''), 
and (b)  the event fraction for which the quark is moving in the same direction as the $W'$. }
\label{sysY.FIG}
\end{figure}

The solid histograms (labeled by ``total'') in the two panels of Fig.~\ref{yWp.FIG} show the reconstructed 
rapidity distribution of 
the partonic c.~m.~frame for $W'_{R}$ and $W'_{L}$ respectively. In comparison, the dashed histograms (labeled
by ``correct direction'') indicate only those events in which the quark is truly along the $W'$ direction
of motion. 
We see that indeed for $| y_{W'} | >1.3$, the kinematically reconstructed $W'$ moves mostly along the quark direction.
To be more quantitative,  Fig. \ref{yWpfrac.FIG} shows the fraction of the events the quark and the reconstructed 
$\wpri$ are moving in the same direction versus the $\wpri$ rapidity. 
Thus if we apply an appropriate lower bound on the $\wpri$ reconstructed rapidity (say $y_{W'}>0.8$), 
the direction of the quark is mostly in the direction of the $\wpri$ and hence, the top-quark
scattering angle $\theta_t$ can be properly defined experimentally.  
This observation, as proposed and studied earlier in \cite{Langacker:1984dc}, 
has already been carefully examined in Ref.~\cite{Dittmar:1996my} 
for neutral current di-lepton events from a heavy $Z'$ at the LHC.
In the above treatment, we have imposed the cuts in Eq.~(\ref{cuts2}) to ensure the $\wpri$ is nearly on-shell.
  We did not smear the energy nor require a $b$-tagging.

\begin{figure}[tb]
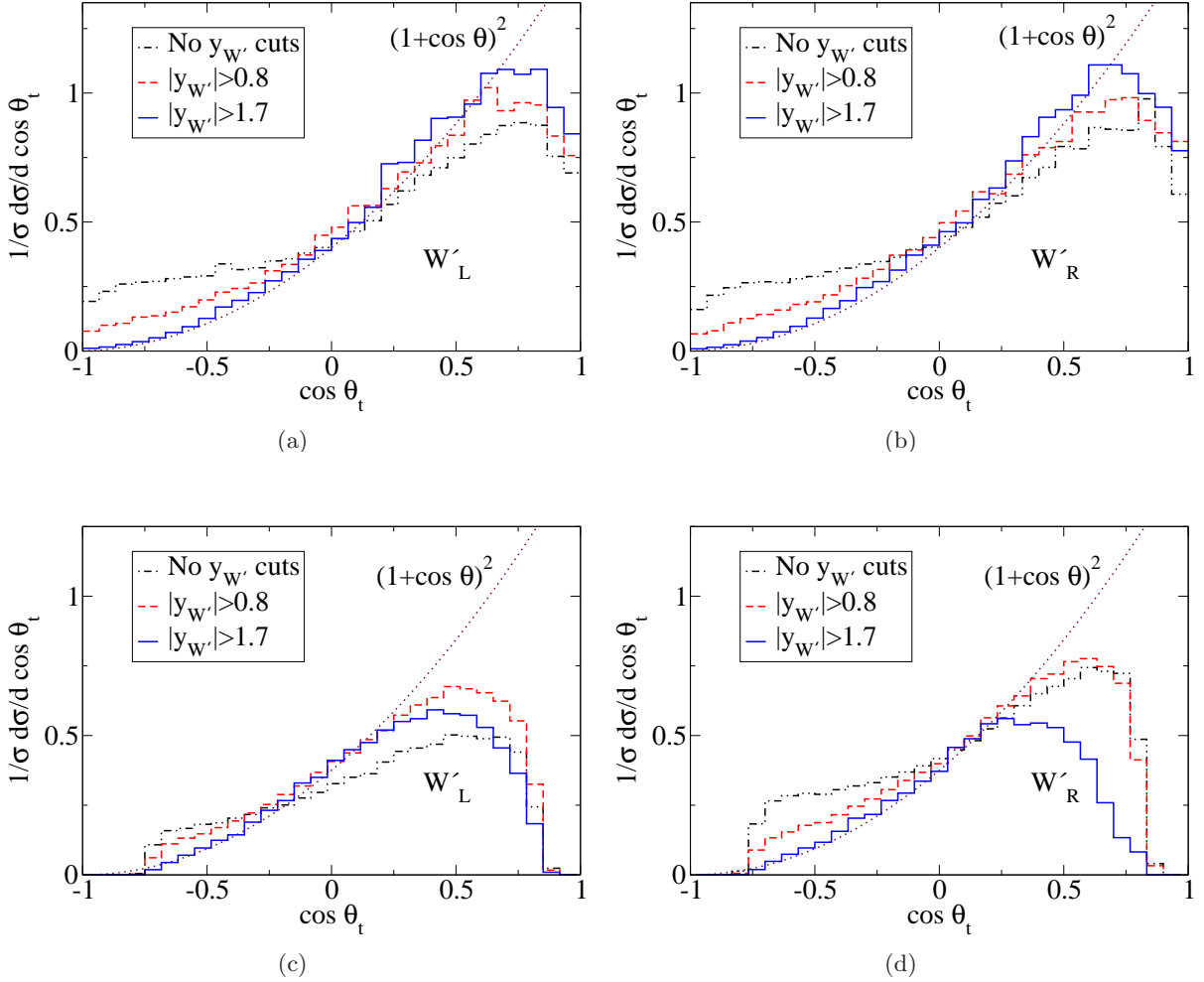

\centering
\subfigure[]{
	\includegraphics[width=0.47\textwidth]{wctrecLHnorap.eps}
	\label{wctrecLHorap.FIG}
         }
\subfigure[]{
	\includegraphics[width=0.47\textwidth]{wctrecRHnorap.eps}
	\label{wctrecRHnorap.FIG}
	}\vspace{.22in}\\
\subfigure[]{
	\includegraphics[width=0.47\textwidth]{wctrecLHalltag.eps}
	\label{wctrecLH.FIG}
         }
\subfigure[]{
	\includegraphics[width=0.47\textwidth]{wctrecRHalltag.eps}
	\label{wctrecRH.FIG}
	}\\
\caption{The angular distribution of the top-quark in the reconstructed c.m.~frame in 
$pp\to W'\to t\bar{b} $ production at the LHC for $M_{W'}=1$ TeV,
with different rapidity cuts,  with (a) and (c) for $W'_L$ and (b) and (d) for $W'_R$.
} 
\label{tcth.FIG}
\end{figure}

To illustrate the validity of the reconstruction method and the effects of kinematical cuts, Fig.~\ref{tcth.FIG} shows the reconstructed top-quark angular distribution with various kinematical cuts and normalizations.  The histograms in Figs.~\ref{tcth.FIG}~(a) and (b) show the reconstructed angular distribution of the top-quark 
for $W'_L$ and $W'_R$, respectively, in the reconstructed c.m.~frame, with various cuts on the partonic c.m.~frame rapidity as listed in the figures and all of the cuts from Eqs.~(\ref{cuts1}), (\ref{cuts3}),
and (\ref{cuts2}) except the final state particle rapidity cuts in Eq.~(\ref{cuts1}).  Each angular distribution has been normalized to one.
The dotted curves present the simple form of  $(1+\cos\theta_{t})^{2}$, 
as expected for a pure (L or R) chiral coupling. If the direction of the quark is misidentified, the top will be identified as going backward (forward) instead of forward (backward) in the partonic c.m.~frame.  Without the $W'$ rapidity cut (the dash-dotted histograms) misidentification is frequent and populates the backward region in the angular distribution. The solid histograms are with the
most stringent $W'$ rapidity cut $|y_{W'}| >1.7$, which follow the dotted curves most faithfully, indicating 
the better  choice of the quark momentum direction.  

Figures~\ref{tcth.FIG}~(c) and (d) show the top-quark angular distribution in the reconstructed c.m.~frame for the $W'_L$ and $W'_R$, respectively, with all the cuts from Eqs.~(\ref{cuts1}),~(\ref{cuts3}),~(\ref{cuts2}), and (\ref{cuts4}) and various cuts on the partonic c.m.~frame rapidity as listed in the figures.  To illustrate the effects of the additional kinematical cuts, instead of being normalized to one, each angular distribution is now normalized using the cross section that was employed in the normalization of the corresponding angular distribution in Figs.~\ref{tcth.FIG}~(a) and (b).  Again, the solid histograms are with the most stringent $W'$ rapidity cut and still follow the $(1+\cos\theta_{t})^{2}$ curve most faithfully for $\cos\theta_t\lesssim 0.25$ .  For all $W'$ rapidity cuts, the cut of Eq.~(\ref{cuts4}) tends to eliminate events where $\cos\theta_t\approx\pm1$.  The additional deficit or distortion in the high rapidity region 
 for $\cos\theta_t\gtrsim0.25$ is due to the final state rapidity cuts of Eq.~(\ref{cuts1}). 

\begin{figure}[tb]
\centering
\subfigure[]{
	\includegraphics[width=0.23\textwidth,clip=true]{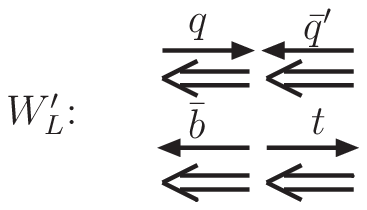}
       \label{TspinLH.FIG}
        }\hspace{20pt}
\subfigure[]{
       \includegraphics[width=0.23\textwidth,clip=true]{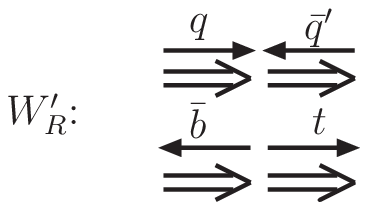}
%       \label{TspinRH.FIG}
        }\vspace{0.15in}
\caption{The helicity and spin correlations between 
the initial-state $q \bar{q}'$ and final state $t \bar{b}$ for (a) and (b) with $\wpri_L$ and $\wpri_R$ 
respectively.  The single arrowed lines show the momentum of the particle, and the double arrowed lines show the spin.
} 
\label{tspin.FIG}
\end{figure}

This well-known distribution in Fig.~\ref{tcth.FIG} can be easily understood based on the argument of spin correlations,
as shown in 
Figs.~\ref{tspin.FIG}~(a) and (b), for the $\wpri_L$ and $\wpri_R$ cases, respectively.  Once again, 
the single arrow lines represent the particle momenta and the double arrow lines the particle helicities. 
For our process, the top-quark is very 
energetic and its helicity state largely coincides with a fixed chirality state. 
A $\wpri_L$ ($\wpri_R$) should couple to left-handed (right-handed) particles and right-handed (left-handed) antiparticles.  
Since the top-quark spin is preferentially in the same direction as the spin of the initial-state quark, 
the top-quark momentum will preferentially be in the same direction as the initial-state quark momentum. 
This argument works for both the $\wpri_L$ and the $\wpri_R$ cases. 
This feature can be confirmed from our analytical expression in Eq.~(\ref{eqn:wct}), where the symmetry
under $g_{L}^{} \leftrightarrow g_{R}^{}$ is evident. 

For a more general case, $g_{L}^{}$ and $g_{R}^{}$ may be both nonzero and nonequal.
The relevant 
observable to determine chiral couplings is the forward-backward asymmetry, defined by
\beq
A_{FB}=\frac{\sigma(\cos\theta_t>0)-\sigma(\cos\theta_t<0)}{\sigma(\cos\theta_t>0)+\sigma(\cos\theta_t<0)},
\label{Eq:AFB}
\eeq
which reflects the parity property of the interaction. 
Using Eq.~(\ref{eqn:wct}),% under the narrow width approximation for the $\wpri$, 
we find that in the partonic c.m.~frame 
\beq
\hat{A}^{qq'}_{FB}=\frac{3}{4\left(1+x^2_t/2\right)}\frac{\left({g^{tb}_R}^2-{g^{tb}_L}^2\right)}{\left({g^{tb}_R}^2+{g^{tb}_L}^2\right)}\frac{\left({g^{qq'}_R}^2-{g^{qq'}_L}^2\right)}{\left({g^{qq'}_R}^2+{g^{qq'}_L}^2\right)}\label{afbpart.EQ}.
\eeq
Convolving with the parton distribution functions and using the narrow width approximation for the $\wpri$,  the hadronic forward-backward asymmetry is
\beq
A_{FB}=\frac{3}{4\left(1+x^2_t/2\right)}\frac{\left({g^{tb}_R}^2-{g^{tb}_L}^2\right)}{\left({g^{tb}_R}^2+{g^{tb}_L}^2\right)}\frac{\sum_{qq'}\left({g^{qq'}_R}^2-{g^{qq'}_L}^2\right)(q\otimes q')(\tau_0)}{\sum_{qq'}\left({g^{qq'}_R}^2+{g^{qq'}_L}^2\right)(q\otimes q')(\tau_0)},\label{afb.EQ}
\eeq
where $\tau_0=M^2_{\wpri}/S$, and the $\otimes$ indicates a convolution defined by
\beq
(q\otimes q')(x)=\int^1_x\frac{dz}{z}q(z)q'\left(\frac{x}{z}\right).
\eeq
Since the narrow width approximation was used, $x_t$ is now $m_t/M_{W'}$.
For a $\wpri$ mass of $1$~TeV and purely left-handed or right-handed couplings to SM fermions, the theoretical value of the partonic and hadronic forward-backward asymmetry is $0.74$.

Table~\ref{afb.TAB}  presents the total cross section and the predicted $A_{FB}$ for pure L and R chiral couplings   
with $M_{W'}=1$ TeV for various values of cutoff on $y^{}_{W'}$.  
The other cuts applied are given in Eqs.~(\ref{cuts1}),~(\ref{cuts3}),~(\ref{cuts2}), and (\ref{cuts4}).
As noted in Eq.~(\ref{eqn:wct}), forward-backward asymmetry for $W'_L$ and $W'_R$ are largely the same, 
with the differences due to how the kinematical cuts affect the two cases differently. 
Also, as the cut on the c.m.~frame rapidity increases, reconstructed and true forward-backward asymmetries converge.   
The forward-backward asymmetry is lowered from the maximum theoretical value because of the reduction
in the forward region.
As seen explicitly in  Eqs.~(\ref{eqn:wct}), (\ref{afbpart.EQ}), and (\ref{afb.EQ}) and demonstrated in this Table, 
the top-quark angular distribution and forward-backward asymmetry cannot distinguish between 
the $\wpri_L$ and $\wpri_R$ cases.  Different values from the prediction in this table may help probe the
mixed nature of the chiral couplings. 

\begin{table}[tb]
\caption{Cross section for $pp\to W'\to t\bar{b}\to b\bar{b}\ \ell^+\nu_\ell$ production at the LHC for $M_{W'}=1$ TeV 
and forward-backward asymmetry of the top-quark in the partonic c.m.~frame for various values of cutoff on $y^{}_{W'}$.
\label{afb.TAB}}
\begin{center}
\begin{tabular}{|c|c|c|c|c|}  \hline
%$y_{cut}$& \multicolumn{2}{|c|}{$W'_L$} &\multicolumn{2}{|c|}{$W'_R$}   \\ \hline \hline
$\sigma$ (fb) & \multicolumn{2}{|c|}{$W'_L$} &\multicolumn{2}{|c|}{$W'_R$}   \\ \hline \hline
$y_{cut}>0$  & \multicolumn{2}{|c|}{110} &\multicolumn{2}{|c|}{150}   \\ \hline 
$0.8$  & \multicolumn{2}{|c|}{55} &\multicolumn{2}{|c|}{78}   \\ \hline 
$1.2$  & \multicolumn{2}{|c|}{30} &\multicolumn{2}{|c|}{42}   \\ \hline 
$1.7$  & \multicolumn{2}{|c|}{7.6} &\multicolumn{2}{|c|}{9.9}   \\ \hline \hline
$A_{FB}$ & True kinematics & Reconstructed & True kinematics & Reconstructed \\ \hline
$y_{cut}>0$  & 0.61 & 0.32 & 0.61 & 0.32 \\ \hline
%$0$   & 0.70     &0.34           & 0.65        &0.34  \\ \hline
$0.8$ & 0.59 & 0.45 & 0.58 & 0.45 \\ \hline
%$0.8$ & 0.65     &0.48           & 0.62        &0.45  \\ \hline
%y_{cut}=1.0$ & 0.51        &   0.47              \\ \hline
$1.2$ & 0.56 & 0.49 & 0.54 & 0.48 \\ \hline
%$1.2$ & 0.60     &0.51           &0.56         &0.47  \\ \hline
%$y_{cut}=1.4$ & 0.52        &   0.46              \\ \hline
$1.7$ & 0.47 & 0.45 & 0.40 & 0.39 \\ \hline
%$1.7$ & 0.50     &0.47           &0.39         &0.37  \\ \hline
\end{tabular}
\end{center}
\end{table}
%

%%%%%%%%%%%%%%%%%%%%%%%%%%%%%%%%%%%
\subsection{$W'$ Chiral Couplings From Top Spin Correlation With\\ Lepton Angle}
\label{Wptspin.SEC}

The chirality of the $W'$ coupling to SM fermions is best encoded in the polarization of the top-quark.
Following the arguments presented in \cite{Frere:1990qm,Tait:2000sh}, 
one may utilize the spin correlation of the top-quark. As already discussed, 
Fig.~\ref{Lspin.FIG} shows the full spin correlations of the relevant particles in the initial and final states for intuitive
understanding. 
Under the narrow width approximation of the top-quark,  the production of $t\bar{b}$ and decay 
$t\to b\ell^+\nu_\ell$ can be factorized.  We start by considering the polarized top-quark production
\beq
q'(p_1)\ \bar{q}(p_2) \to W' \to t(k_1,s_t)\  \bar{b}(k_2),
\label{qq2Wp2tb.EQ}
\eeq
%where $s_t$ is the top spin vector.  
The top spin vector has the properties
\bea
s_t^2\, =\, -1,\  k_1\cdot s_t \, = \, 0, \quad \text{and}~~~
s_t^{\mu} \, =\, \left( \frac{\vec{k}_1\cdot \hat{s}_t}{m_t}, \,
\hat{s}_t\,+\,\frac{\vec{k}_1 (\vec{k}_1\cdot \hat{s}_t )}{m_t\ (m_t +E_1)}
\right),
\eea
where $E_1$ and $\vec{k}_1$ are the energy and the three-momentum of the top-quark 
in the partonic c.m.~frame and $\hat{s}_t$ the top spin three-vector in its rest frame. 
The cross section for the process in Eq.~(\ref{qq2Wp2tb.EQ}) is of the following form
\beq
d\sigma(s_t)\, = \, 
\frac{1}{2\hat{s}}\, \frac{1}{4}\, dPS_2\, |{\cal M}|^2,\quad
|{\cal M}|^2 \, =\,\frac{g_2^4}{4}\bigg{(} A_0 \, + \, B_{\mu} s_t^{\mu}\bigg{)},
\eeq
where $dPS_2$ is the 2-particle phase space and $|{\cal M}|^2$ is the polarized
invariant amplitude squared.
%We compute analytically $|{\cal M}|^2$ and 
We find 
\bea
A_0 &=&\frac{8}{(\hat{s}-M_{W'}^2)^2+\Gamma^2_{W'} M^2_{W'}}
\, \Big\{
\left[ g_L^{tb\ 2}\, g_L^{qq'\ 2} \ + \
g_R^{tb\ 2}\ g_R^{qq'\ 2}\right]\ (k_1\cdot p_2)\ (k_2\cdot p_1) \nonumber \\
&&\, + \, \left[g_L^{qq'\ 2}\ g_R^{tb\ 2}\ +\ g_L^{tb\ 2}\ g_R^{qq' 2}\right]
(k_1\cdot p_1) (k_2\cdot p_2)
-\frac{\hat{s}}{2} m_b m_t \, \Big[g_L^{qq'\ 2}+g_{R}^{qq'\ 2}\Big] g_L^{tb} g_R^{tb}
\Big\},\label{eq:A0}\\
B_{\mu} &=&\frac{8\ m_t}{(\hat{s}-M_{W'}^2)^2+\Gamma^2_{W'} M^2_{W'}}
\, \Big\{
 \left[ g_L^{qq'\ 2}\ g_R^{tb\ 2}\ -\ g_L^{tb\ 2}\ g_R^{qq'\ 2}\right]
(k_2\cdot p_2)\ p_{1\mu}  \nonumber \\
&&\, -\, \frac{m_b}{m_t}\, g_L^{tb}\, g_R^{tb}\, (g_L^{qq'\ 2}-g_{R}^{qq'\ 2})\, \Big[
(k_1\cdot p_2)\, p_{1\mu}\, -\, (k_1\cdot p_1)\, p_{2\mu}
\Big]\nonumber\\
&& \, + \,
\left[ g_R^{tb\ 2}\ g_R^{qq'\ 2}\ -\ g_L^{tb\ 2}\ g_L^{qq'\ 2}\right ]\
(k_2\cdot p_1)\ p_{2\mu} \,
\Big\}. \label{eq:Bmu} 
\eea
The $B_\mu s^\mu_t$ term contains information about whether the $\wpri$ has purely right-handed or left-handed chiral couplings.  By averaging over the top-quark spin we reproduce Eq.~(\ref{eqn:wct}) but lose the ability to determine whether the $\wpri$ is right-handed or left-handed.  To discriminate between the right-handed and left-handed cases we need an observable that is dependent on the $B_\mu s^\mu_t$ term.

We now consider the polarized top-quark decay. 
The terms related to $B_{\mu} s_t^{\mu}$ will contribute to the spin
observables $\left<\hat{s}_t \cdot \hat{a}\right>$ where $\hat{a}$ is a 
suitably chosen spin-quantization axis. 
Obviously, this observable 
will have different sign for pure left-handed and right-handed interactions.
In terms of an angle ($\theta_\ell$) between the direction of the charged lepton 
and the direction $\hat{a}$ \cite{Bernreuther:2001rq} in the top rest frame, the
angular distribution for a polarized top decay can be written as
\beq
\frac{d\Gamma_{s_{t}}}{d\cos\theta_\ell} \, =\,\frac{1}{2}\,\sigma\, \Big\{
1\,+\, 2\, A \ \cos \theta_\ell  \Big\},
\label{dsdxCdef.EQ}
\eeq
where $A$ can be related to an observed  forward-backward asymmetry similarly to Eq.~(\ref{Eq:AFB}). 
While the direction $\hat{a}$ can be chosen freely, different choices can change the magnitude of $A$, thus called an ``analyzing
power''. 
The simplest choice of $\theta_\ell$ can be taken to be the angle between the lepton and the $b$-jet.
Specifically for the process under consideration, we may take the $b$-jet that is closer to the lepton 
in the lab frame, since this one is most likely to originate from the top decay.  With that choice of $\hat{a}$ one finds that the analyzing
power is  $A=0.36\ (0.35)$ for the $\wpri_L$ ($\wpri_R$), without smearing or cuts, which 
does not clearly differentiate between the two cases L and R. This is due to the kinematics that 
the top-quark is highly boosted for both $\wpri_L$ and $\wpri_R$, 
and also due to the relative kinematics between the $b$ and lepton being the same 
for both the $\wpri_L$ and $\wpri_R$ cases as can be inferred from Fig.~\ref{Lspin.FIG}. 
A similar effect happens when $\theta_\ell$ is the opening angle between the lepton and the $b$-jet from the $\wpri$ decay.  
Hence, we need to reconstruct the system and define a suitable direction $\hat{a}$.

At high energies, 
a natural choice of $\hat{a}$ is the top-quark three-momentum direction in the $t\bar{b}$ c.m.~frame.  Since the top is nearly in a helicity state, $\left<\hat{s}_t \cdot \hat{k}_1\right>$ should be large and increase our sensitivity to the chiral couplings. 
With this choice and using the narrow width approximation for the top-quark and $W$ decays, 
the lepton angular distribution in the top-quark rest frame is
\bea
\frac{d\hat\sigma(q\bar q'\to W'\to t\bar b\to b\bar b \ell\nu)}{d\cos\theta_\ell}  
= \frac{1}{2}\hat\sigma_0\bigg{(}1+\frac{{g^{tb}_R}^2-{g^{tb}_L}^2}{{g^{tb}_R}^2+{g^{tb}_L}^2}\frac{\hat s-m^2_t/2}{\hat s+m^2_t/2}\cos\theta_\ell\bigg{)},
\label{dsigdx.EQ}
\eea
where
\bea
\hat\sigma_0&=& \frac{{g_2}^8}{3\cdot 2^{15}\pi^3} \frac{m_t M_W}{\Gamma_t\Gamma_W}\
\big{(}{g^{qq'}_L}^2+{g^{qq'}_R}^2\big{)}\big{(}{g^{tb}_L}^2+{g^{tb}_R}^2\big{)} \big{(}x^2_{wt}-1-\log x^2_{wt}\big{)}
\nonumber\\
&\times&\big{(}1-x^2_t\big{)}^2\bigg{(}1+\frac{x^2_t}{2}\bigg{)}\frac{\hat s}{(\hat{s}-M^2_{W'})^2+\Gamma^2_{W'}M^2_{W'}},
\label{sig0full.EQ}
\eea
with $x_{wt}=M_W^{}/m_t$ and $x_t=m_t/\sqrt{\hat s}$.  
Some technical details of this calculation are  given in Appendix~\ref{LepAngDist.APP}.  
For an on-shell $W'$,  we can read the analyzing power as 
\beq
A \, = \, \frac{1}{2}\left(\frac{{g^{tb}_R}^2-{g^{tb}_L}^2}{{g^{tb}_R}^2+{g^{tb}_L}^2}\right)
\left(\frac{M^2_{W'}-m^2_t/2}{M^2_{W'}+m^2_t/2}\right) \ .
\label{eq:C.FORM}
\eeq
It is evident that $A$ has opposite sign for pure left-handed and pure right-handed cases,
and lies between $-1/2$ and $1/2$ for a L-R mixed $W'$.
The observables in Eqs.(\ref{dsigdx.EQ})~and~(\ref{eq:C.FORM}) are sensitive to how the top is polarized which
in turn only depends on $g^{tb}_{L,R}$ and not on $g^{qq'}_{L,R}$; the kinematics of 
top production, for instance the $\theta_t$ distribution discussed in Sec.~\ref{WpriAC.SEC}.A,%\ref{TopAngDist.SEC},
 however, are sensitive to the latter coupling as well.

\begin{figure}[tb]
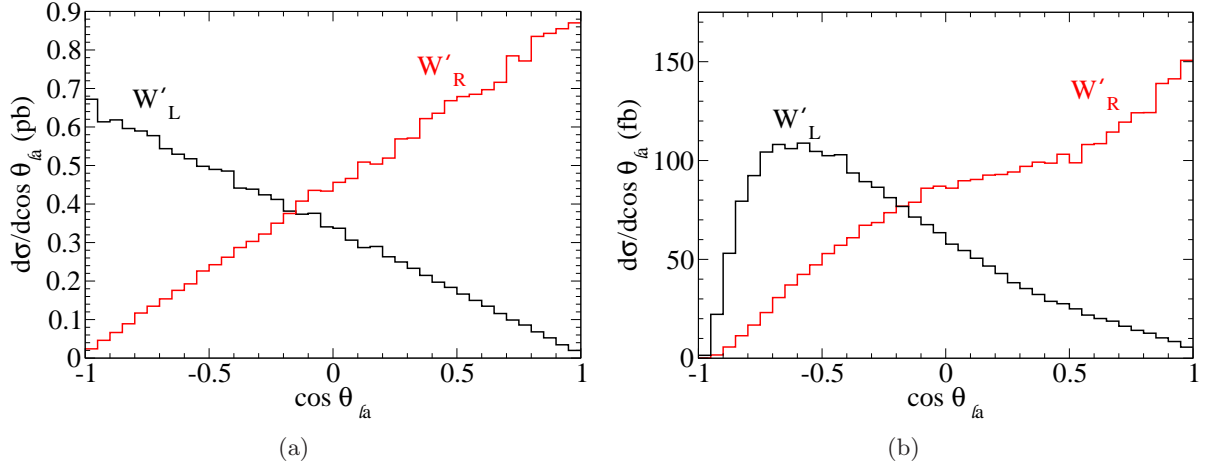

\begin{center}
\subfigure[]{
	\includegraphics[width=0.465\textwidth]{ctlepNCNS.eps}
	\label{ctlepNCNS.FIG}
	}
\subfigure[]{
        \includegraphics[width=0.47\textwidth]{ctlepalltag.eps}
        \label{ctlepall.FIG}
        }
\caption{
The angular distribution of the charged lepton in 
$pp\to W'\to t\bar{b}\to b\bar{b}\ \ell^+\nu_\ell$ production at the LHC for $M_{W'}=1$ TeV 
in the top-quark rest-frame with respect to a spin quantization direction $\hat{a}$ 
taken to be the top direction in the c.m.~frame,
for 
(a) without smearing or cuts, and (b) with energy smearing and cuts 
in Eqs.~(\ref{cuts1}),(\ref{cuts3}),(\ref{cuts2}),(\ref{cuts4}), and tagging the softest $b$-jet. }
\label{wpriac.FIG}
\end{center}
\end{figure}

Using the reconstructed events as discussed in the previous section, 
we can find the angular distribution of the lepton in the top rest frame. 
The results of this reconstruction are shown in Fig.~\ref{wpriac.FIG} after convolving with the parton distribution functions.  
We see the distinctive distributions for the L or R chiral couplings from Fig.~\ref{wpriac.FIG}(a). 
The charged lepton prefers to move against the direction of the top-quark in the partonic c.m.~frame in the $\wpri_L$ case;
while in the $\wpri_R$ case  the charged lepton prefers to move in the same direction as the top-quark
as can be seen in the illustration in Fig.~\ref{Lspin.FIG}.
We also illustrate the realistic situation with kinematical cuts as in  Fig.~\ref{wpriac.FIG}(b). 
The main effect of the cuts is to flatten out the distributions somewhat.  Around $\cos\theta\approx -1$ the lepton becomes soft and fails the transverse momentum cut.  Hence the $W_L'$ distribution is affected more by the cuts than the $W_R'$ cut.  In spite of the effects of the cuts, we retain an excellent ability to tell $W_L'$ and $W_R'$ apart.

Using the reconstructed events we can also determine the asymmetric observable $A$.
The results for $A$ are given in Table~\ref{Aasym.TAB} for the signal of $W'_{L}$ and $W'_{R}$
with and without including the SM $W$ contribution. 
To demonstrate the realistic kinematical effects, we give the asymmetries with consecutive cuts in the table. 
Once all the cuts have been applied we still obtain a very good determination of the chirality of the $W'$. 
\begin{table}[tb]
\caption{Forward-backward asymmetry of the charged lepton in $pp\to t\bar{b}\to b\bar{b}\ell^+\nu_\ell$ for $\ell=e^+$ or $\mu^+$
at the LHC for $M_{W'}=1$TeV with and without the SM $W$ contribution.
\label{Aasym.TAB}}
\begin{center}
\begin{tabular}{|l|c|c|c|c|}  \hline
$A$                  & $W+W'_L$  & $W+W'_R$ & $W'_L$  & $W'_R$ \\
 \hline
No Cuts or smearing  & $-0.42$        &   $0.17$       &$-0.48$  &$0.48$  \\ \hline
No Cuts              & $-0.42$        &   $0.15$       &$-0.49$  &$0.45$  \\ \hline
Cuts Eqs.(\ref{cuts1})   & $-0.48$        &  $0.24$        &$-0.51$  &$0.37$  \\ \hline
~~+Eq.(\ref{cuts2})&$-0.49$         & $0.39$         &$-0.49$  & $0.40$ \\ \hline
~~+Eq.(\ref{cuts3})& $-0.53$        & $0.36$         &$-0.53$  & $0.37$ \\ \hline
~~+Eq.~(\ref{cuts4}) \& tagging 1 $b$-jet &$-0.48$ & $0.40$ &$-0.48$ &$0.40$ \\ \hline
\end{tabular}
\end{center}
\end{table}

%%%%%%%%%%%%%%%%%%%%%%%%%%%%%%
\subsection{$W'$ Chiral Couplings From Transverse Momentum Distributions}
\label{W'chilpT.SEC}
As discussed already in Sec.~3.2, the $p_T$ distributions also convey information on the
$W'$ chirality as shown in Fig.~\ref{ptdist.FIG} due to their spin correlations. 
The charged lepton $p_T$ in the case of $W_R'$ is harder  
than that in $W'_L$. This can be understood from angular-momentum conservation;
for the former the charged lepton moves preferentially in
the same direction as the top thereby boosting the lepton $p_T$, 
while in the latter it is against this making the lepton $p_T$ softer. 
The $p_T$ of the $b$-jet from top decay on the other hand has the opposite relationship.
Although the $W'$ chirality affects the lepton $p_T$ (and the softer-jet $p_T$), 
measuring it will require discerning
this rather subtle modification which will be limited by statistics.

%%%%%%%%%%%%%%%%%%%%%%%%%%%%%%%%%%%%%
\subsection{$W'$ Chiral Couplings From Other Variables}
In this section we explore some other variables that can be used to fix the chirality of the
$\wpri$ couplings. 

\begin{figure}[tb]
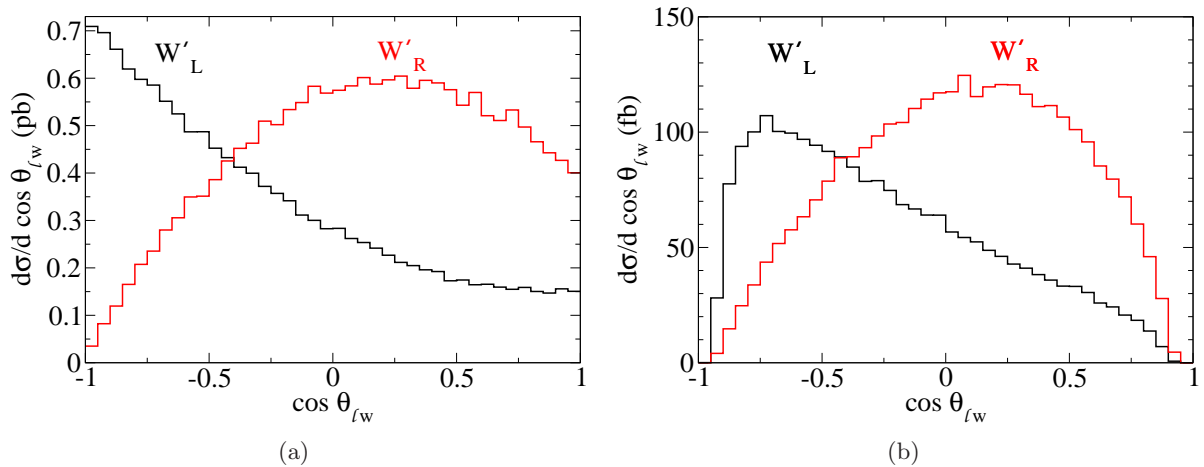

\begin{center}
\vspace{0.2in}
\subfigure[]{
	\includegraphics[width=0.465\textwidth]{costhlwNCNS.eps}
	\label{costhlwNCNS.FIG}
	}
\subfigure[]{
        \includegraphics[width=0.47\textwidth]{costhlwalltag.eps}
        \label{costhlwall.FIG}
        }
\caption{
The angular distribution of the charged lepton in 
$pp\to W'\to t\bar{b}\to b\bar{b}\ \ell^+\nu_\ell$ production at the LHC for $M_{W'}=1$ TeV 
in the $W$ rest-frame with respect to the $W$ moving direction 
for (a) without smearing or cuts, and (b) with energy smearing and cuts 
Eqs.~(\ref{cuts1}),(\ref{cuts3}),(\ref{cuts2}),(\ref{cuts4}), and tagging the softest jet.
\label{wpriac2.FIG}
}
\end{center}
\end{figure}

Because of the spin correlations in the production and decay chain, the $W$ is also polarized 
in a definite manner. 
For example, we consider the variable $\cos{\theta_{\ell\hat{w}}}$, 
the angle the lepton makes in the $W$ rest frame relative to the direction ($\hat{w}$) 
of the $W$ in the lab frame with our fully reconstructed events.
In Fig.~\ref{wpriac2.FIG} we show its differential distribution for pure left- and right-chiral
cases.
As can be seen in the figure, we have an ability to determine the $W'$ chirality
using this observable, but for a reason not as direct as in the previous subsections.
From Fig.~\ref{Lspin.FIG} we see that the $\cos{\theta_\ell}$ distribution with respect to 
the $W$ direction is the same for both $\wpri_L$ and $\wpri_R$, 
but one has to keep in mind that in this figure the $W$ direction 
is as seen in the top rest frame.  
However, the boost of the top plus the selection of the large system rapidity by the $|y_{\wpri}|$
cut carries the $W$ in the quark direction in the lab-frame
effectively flipping the $W$ momentum arrow in Fig.~\ref{Lspin.FIG}~(d). 
It then becomes clear why the $\cos{\theta_{\ell\hat{w}}}$ distribution ends up distinguishing between 
$\wpri_L$ and $\wpri_R$ as seen in Fig.~\ref{wpriac2.FIG}. 
The kinematics we have here ends up being favorable for this variable to differentiate between 
the chiralities and may not be as general a method as those presented in the previous 
subsections. 

In our work, we have focused on angular correlations of the lepton since it is experimentally clean, 
but one can consider angular correlations of jet observables also. 
For instance, the angular distribution with respect to the top direction 
of the (softer) $b$-jet coming from top decay in the top rest frame  
also analyzes the top polarization and thus can be used to extract 
$g^{tb}_{L,R}$. Another example is the angular distribution of the harder $b$-jet 
with respect to the quark direction which 
carries the same information as the top angular distribution given in Eq.~(\ref{eqn:wct}).  
Rather than analyzing the variables in the top rest frame as we have chosen to do, 
one can consider observables in the lab frame or with the event partially 
reconstructed that are sensitive to the chirality as discussed elsewhere~\cite{OtherTopPol}.

%%%%%%%%%%%%%%%%%%%%%%%%%%%%%%%%%%%%%%%%%%%%%%%%%%%%%%%%%%%%%%%%%%%%%%%%%%%%%%%%%%%%%
\section{DISCUSSION AND CONCLUSIONS}
\label{DiscConcl.SEC}
Many beyond the SM theories predict new heavy gauge bosons.  If the masses of these new resonant states are not more than a few TeV and their couplings to SM fermions are not too suppressed, the LHC has a good chance of discovering these new resonant states.  After discovery, it will
become imperative to determine the properties of the new gauge boson, such as spin, mass, and couplings.  Of them all, the critically 
important feature would be their chiral couplings to SM fermions.

In this paper we focused on measuring the chiral couplings between SM fermions and a new heavy, charged gauge boson ($\wpri$).  Specifically, the chiral couplings of $\wpri~t~b$ is determined through the process $pp\rightarrow \wpri\rightarrow t\bar b\rightarrow \ell^+\nu_\ell b\bar b$ by fully reconstructing the event and forming various angular distributions.

We have demonstrated that it is possible to reconstruct the forward-backward asymmetry of the top-quark in the partonic c.m.~frame.  However, this forward-backward asymmetry is unable to distinguish between purely right-handed and purely left-handed chiral couplings.  The chiral nature of the $\wpri~t~b$ coupling is reflected in the top-quark polarization.  A good diagnostic of the 
top-quark polarization is the forward-backward asymmetry of the lepton.
We have shown that by measuring the lepton's forward-backward asymmetry in the 
top-quark's rest frame one can determine the $\wpri~t~b$ chiral couplings.  
Once the $\wpri~t~b$ chiral couplings are determined, 
the top-quark's forward-backward asymmetry in the partonic c.m.~frame  can be used to determine the 
chiral couplings of the $\wpri$ with the initial-state quarks.

The reconstruction method described in this paper is quite general.  
The procedure to determine the $\wpri$ chiral couplings to SM fermions described in this paper 
could be adapted to gauge bosons decaying to new heavy fermions.

\vspace*{0.3cm}

\noindent {\it Acknowledgments:} We thank R.~Godbole for valuable discussions, and P.~Langacker 
for a discussion on electroweak precision bounds. 
TH and IL were supported in part by the US DOE under contract No.~DE-FG02-95ER40896, 
SG was supported in part by the DOE grant DE-AC02-98CH10886 (BNL) at the 
Brookhaven National Laboratory where a part of this work was carried out 
and in part by the Australian Research Council, 
and ZS was supported in part by NSFC and Natural Science Foundation of 
Shandong Province (JQ200902). 
We thank KITP Santa Barbara for hospitality where this work was initiated, 
supported in part by the National Science Foundation under Grant No. PHY05-51164.
TH also acknowledges the hospitality of Aspen Center for Physics during the final stage of this project. 

%%%%%%%%%%%%%%%%%%%%%%%%%%%%%%%%%%%%%%%%%%%%%%%%
\appendix

\section{LEPTON ANGULAR DISTRIBUTION}
\label{LepAngDist.APP}
Here we present details on deriving the lepton angular distribution in the 
top rest-frame.  First, the partonic cross section is given by the usual formula
\begin{eqnarray}
d\hat\sigma=\frac{1}{2\hat s}\frac{1}{4N^2_C}\sum|\mathcal{M}|^2dPS_4,
\label{eq:dsig1}
\end{eqnarray}
where $N_C=3$ is the number of colors, the sum is over the external particle spins and colors, $\mathcal{M}$ is the matrix element, and $dPS_4$ is the four body phase space of the final state lepton, neutrino, and two bottom quarks.  Using the narrow width approximation for the top-quark and SM W, the sum over the matrix element squared is found to be
\begin{eqnarray}
\sum|\mathcal{M}|^2&=&\frac{8N^2_C{g_2}^8\pi^2}{\Gamma_t m_t\Gamma_WM_W}\frac{\delta(k^2_1-m^2_t)\delta(q^2_2-M^2_W)}{(\hat{s}-M^2_{W'})^2+\Gamma^2_{W'}M^2_{W'}}k_3\cdot k_\nu\nonumber\\
&\times&\bigg{\{}~{g^{qq'}_L}^2{g^{tb}_L}^2p_1\cdot k_{2}(2p_{2}\cdot k_1 k_1\cdot k_\ell -p_{2}\cdot k_\ell  m^2_t) 
+{g^{qq'}_L}^2{g^{tb}_R}^2p_{2}\cdot k_2 p_1\cdot k_\ell  m^2_t\nonumber\\
&+&{g^{qq'}_R}^2{g^{tb}_L}^2p_{2}\cdot k_2(2p_1\cdot k_1 k_1\cdot k_\ell  -p_1\cdot k_\ell  m^2_t) 
+{g^{qq'}_R}^2{g^{tb}_R}^2p_{1}\cdot k_2 p_{2}\cdot k_\ell  m^2_t\bigg{\}}, \nonumber
\end{eqnarray}
where  the momenta are as labeled in Fig.~\ref{qq2Wp2tb.FIG} and $m_b=0$.  To evaluate the four body phase space we use the phase space recursion formula:
\begin{eqnarray}
dPS_4=dPS_2(k_1,k_2)dPS_3(k_3,k_\ell ,k_\nu)\frac{dk^2_1}{2\pi}.
\end{eqnarray}

The goal is to obtain the lepton angular distribution in the top-quark's rest frame with the $z$-direction defined as the top-quark's direction in the partonic c.m.~frame.  Hence, the matrix element squared needs to be evaluated in the top-quark's rest frame.  Since the momenta $k_1$ and $k_2$ are fixed in the top-quark's rest frame, they cannot be integrated over in the phase space $dPS_2(k_1,k_2)$.  We will thus evaluate the phase space $dPS_2(k_1,k_2)$ in the partonic c.m.~frame and obtain $p_1,p_2,k_1$, and $k_2$ in the top-quark's rest frame in terms of the top-quark angles in the partonic c.m.~frame.  The phase space and matrix element squared can be evaluated in two different frames since they are independently Lorentz invariant.  Once this procedure is complete, we can integrate the matrix element squared evaluated in the top-quark rest frame over the top-quark angles in the partonic c.m.~frame.

First, in the partonic c.m.~frame, the 2-body phase space is given by
\begin{eqnarray}
dPS_2(k_1,k_2)=\frac{1-x^2_t}{2(4\pi)^2}d\cos\theta_1d\phi_1,
\end{eqnarray}
where $x_t=m_t/\sqrt{\hat s}$, $\theta_1$ is the angle the top-quark makes with the initial-state quark in the partonic c.m.~frame, and $\phi_1$ is the top-quark's azimuthal angle in the partonic c.m.~frame.  

To evaluate the momenta in the top-quark's rest frame we start with the initial-state quarks, top-quark, and antibottom quark momenta in the partonic c.m.~frame, rotate the momenta so that the top-quark is in the $z$-direction, and boost the system to the top-quark rest frame.  These operations take us to the top-quark's rest frame with the desired orientation. Assuming that in the partonic c.m.~frame the initial-state quark is moving in the $z$-direction and the top-quark momentum is in the $y-z$ plane, the resulting momenta are found to be
\begin{eqnarray}
p_1&=&\frac{\hat{s}}{4m_t} \bigg{(}1-\cos\theta_1+x^2_t(1+\cos\theta_1),0,-2x_t\sin\theta_1,-(1-\cos\theta_1)+x^2_t(1+\cos\theta_1)\bigg{)},\nonumber\\
p_2&=&\frac{\hat{s}}{4m_t}\bigg{(}1+\cos\theta_1+x^2_t(1-\cos\theta_1),0,~~2x_t\sin\theta_1,-(1+\cos\theta_1)+x^2_t(1-\cos\theta_1)\bigg{)},\nonumber\\
k_1&=&(m_t,0,0,0),\quad {\rm and}\quad k_2=\frac{\hat s}{2m_t}(1-x^2_t)(1,0,0,-1).
\end{eqnarray}
.

Substituting these momenta into the matrix element squared and integrating over the 2-body phase space, the differential cross section becomes
% we obtain the matrix element squared in the top-quark rest frame with the z-axis in the direction of the top-quark in the partonic c.m.~frame in terms of the top-quark angles in the partonic c.m.~frame. 
\begin{eqnarray}
d\hat\sigma&=&\frac{{g_2}^8}{3\cdot 2^7}\frac{\hat s (1-x^2_t)^2}{\Gamma_t\Gamma_WM_W}\frac{k_3\cdot k_\nu E_\ell }{(\hat{s}-M^2_{W'})^2+\Gamma^2_{W'}M^2_{W'}}({g^{qq'}_L}^2+{g^{qq'}_R}^2)({g^{tb}_L}^2+{g^{tb}_R}^2)\nonumber\\
&\times&\bigg{\{}(2+x^2_t)\bigg{[}1+\frac{{g^{tb}_R}^2-{g^{tb}_L}^2}{{g^{tb}_R}^2+{g^{tb}_L}^2}\frac{2-x^2_t}{2+x^2_t}\cos\theta_\ell \bigg{]}+\frac{{g^{qq'}_L}^2-{g^{qq'}_R}^2}{{g^{qq'}_L}^2+{g^{qq'}_R}^2}\frac{3}{4}\pi x_t\sin\phi_\ell \sin\theta_\ell \bigg{\}}\nonumber\\
&\times&\delta(q^2_2-M^2_W)dPS_3(k_3,k_\ell ,k_\nu),
\label{eq:dsig2}
\end{eqnarray}
where $\theta_\ell $ is the charged lepton's polar angle in the top-quark's rest frame with respect to the top-quark's direction in the partonic c.m.~frame, $\phi_\ell $ is the charged lepton's azimuthal angle in the top-quark's rest frame, and $E_\ell $ is the lepton's energy in the top-quark rest frame.

Now, the 3-body phase space in the top-quark's center of mass frame is
\begin{eqnarray}
dPS_3(k_3,k_\ell ,k_\nu)=\frac{1}{(4\pi)^5}2E_3dE_3d\cos\theta_3 d\phi_3d\cos\theta_\ell  d\phi_\ell ,
\end{eqnarray}
where $E_3$ is the bottom quark's energy, $\theta_3$ ($\phi_3$) is the bottom quark's polar (azimuthal) angle in the top-quark rest frame, and the integration over the bottom quark's energy is from $0$ to $m_t/2$.  The delta function in Eq.~(\ref{eq:dsig2}) can be rewritten as
\begin{eqnarray}
\delta(q^2_2-M^2_W)&=&\frac{1}{2m_t}\delta\bigg{(}E_3-\frac{m_t}{2}(1-x^2_{Wt})\bigg{)},
\end{eqnarray}
where $x_{Wt}=M_W/m_t$.  This delta function fixes the bottom quark energy.  Using the conservation of energy and momentum, we can also solve for the lepton and neutrino four momenta.  The relevant quantities for our calculation are
\begin{eqnarray}
E_\ell &=&m_t\frac{x^2_{Wt}}{1+\cos\theta_{3l}+x^2_{Wt}(1-\cos\theta_{3l})}\\
k_3\cdot k_\nu&=&\frac{m^2_t}{2}\frac{(1-x^2_{Wt})(1+\cos\theta_{3l})}{1+\cos\theta_{3l}+x^2_{Wt}(1-\cos\theta_{3l})},
\end{eqnarray}
where $\cos\theta_{3l}=\sin\theta_\ell \sin\theta_3\cos(\phi_\ell -\phi_3)+\cos\theta_\ell \cos\theta_3$ is the cosine of the angle between the lepton and the bottom quark.

Using the above identities and integrating over the bottom quark's energy, the lepton angular distribution becomes
\begin{eqnarray}
\frac{d\hat\sigma}{d\cos\theta_\ell }&=&({g^{qq'}_L}^2+{g^{qq'}_R}^2)({g^{tb}_L}^2+{g^{tb}_R}^2)(1-x^2_t)^2(1-x^2_{Wt})^2\frac{{g_2}^8}{3\cdot 2^{18}\pi^5}\frac{m_tM_W}{\Gamma_t\Gamma_W}\frac{\hat s}{(\hat{s}-M^2_{W'})^2+\Gamma^2_{W'}M^2_{W'}}\nonumber\\
&\times&\bigg{\{}(1+\frac{x^2_t}{2})\bigg{[}1+\frac{{g^{tb}_R}^2-{g^{tb}_L}^2}{{g^{tb}_R}^2+{g^{tb}_L}^2}\frac{2-x^2_t}{2+x^2_t}\cos\theta_\ell \bigg{]}+\frac{{g^{qq'}_L}^2-{g^{qq'}_R}^2}{{g^{qq'}_L}^2+{g^{qq'}_R}^2}\frac{3}{8}\pi x_t\sin\phi_\ell \sin\theta_\ell \bigg{\}}\nonumber\\
&\times&\frac{1+\cos\theta_{3l}}{[1+\cos\theta_{3l}+x^2_{Wt}(1-\cos\theta_{3l})]^2}d\cos\theta_3 d\phi_3d\phi_\ell 
\end{eqnarray}

To simplify the integration over the azimuthal angles, we make the variable transformation
\begin{eqnarray}
\phi_+=\frac{\phi_3+\phi_\ell }{2} , \quad \phi_-=\frac{\phi_3-\phi_\ell }{2}.
\end{eqnarray}
The integration limits for $\phi_-$ and $\phi_+$ are more complicated than the limits for $\phi_\ell$ and $\phi_3$:
\begin{eqnarray}
\int^{2\pi}_0d\phi_3\int^{2\pi}_0d\phi_\ell =2\bigg{\{}\int^0_{-\pi}d\phi_-\int^{2\pi+\phi_-}_{-\phi_-}d\phi_++\int^\pi_0d\phi_-\int^{2\pi-\phi_-}_{\phi_-}d\phi_+\bigg{\}}
\end{eqnarray}
Using the symmetries of the differential cross section to simplify the integration, the lepton angular distribution becomes
\begin{eqnarray}
\frac{d\hat\sigma}{d\cos\theta_\ell }&=&\big{(}{g^{qq'}_L}^2+{g^{qq'}_R}^2\big{)}\big{(}{g^{tb}_L}^2+{g^{tb}_R}^2\big{)}\big{(}1-x^2_t\big{)}^2\big{(}1-x^2_{Wt}\big{)}^2\bigg{(}1+\frac{x^2_t}{2}\bigg{)}\frac{{g_2}^8}{3\cdot 2^{15}\pi^3}\frac{m_tM_W}{\Gamma_t\Gamma_W}\nonumber\\
&\times&\frac{\hat s}{(\hat{s}-M^2_{W'})^2+\Gamma^2_{W'}M^2_{W'}}\bigg{[}1+\frac{{g^{tb}_R}^2-{g^{tb}_L}^2}{{g^{tb}_R}^2+{g^{tb}_L}^2}\frac{2-x^2_t}{2+x^2_t}\cos\theta_\ell \bigg{]}\mathcal{I}_1(\cos\theta_\ell ),
\end{eqnarray}
where
\begin{eqnarray}
\mathcal{I}_1(\cos\theta_\ell )=\frac{1}{\pi^2}\int^{1}_{-1}d\cos\theta_3\int^\pi_0d\phi_-(\pi-\phi_-)\frac{1+\cos\theta_{3l}}{[1+\cos\theta_{3l}+x^2_{Wt}(1-\cos\theta_{3l})]^2},
\end{eqnarray}
and $\cos\theta_{3l}=\sin\theta_\ell \sin\theta_3\cos(2\phi_-)+\cos\theta_\ell \cos\theta_3$.
To find a solution for $\mathcal{I}_1$, the $\phi_-$ integral must be analytically continued into the complex plane and a contour integral completed. The details of this calculation are given in Appendix~\ref{ContInt.APP}.  

Once $\mathcal{I}_1$ is integrated, we have
\begin{eqnarray}
\frac{d\hat\sigma}{d\cos\theta_\ell }&=&\big{(}{g^{qq'}_L}^2+{g^{qq'}_R}^2\big{)}\big{(}{g^{tb}_L}^2+{g^{tb}_R}^2\big{)}\big{(}x^2_{wt}-1-\log x^2_{wt}\big{)}\frac{{g_2}^8}{3\cdot 2^{16}\pi^3}\frac{m_tM_W}{\Gamma_t\Gamma_W}\nonumber\\
&\times&\frac{\hat s}{(\hat{s}-M^2_{W'})^2+\Gamma^2_{W'}M^2_{W'}}\big{(}1-x^2_t\big{)}^2\bigg{(}1+\frac{x^2_t}{2}\bigg{)}\bigg{[}1+\frac{{g^{tb}_R}^2-{g^{tb}_L}^2}{{g^{tb}_R}^2+{g^{tb}_L}^2}\frac{2-x^2_t}{2+x^2_t}\cos\theta_\ell \bigg{]}
\end{eqnarray}
Hence,  at the partonic level, the forward-backward asymmetry of the lepton in the top center of mass frame with the $z$-direction defined as the direction of the top-quark in the partonic c.m.~frame is
\begin{eqnarray}
\hat A \, &=& \, \frac{\hat\sigma(\cos\theta_\ell >0)-\hat\sigma(\cos\theta_\ell <0)}
{\hat\sigma(\cos\theta_\ell >0)+\hat\sigma(\cos\theta_\ell <0)} 
= \frac{1}{2}\left(\frac{{g^{tb}_R}^2-{g^{tb}_L}^2}{{g^{tb}_R}^2+{g^{tb}_L}^2}\right)
\left(\frac{\hat s-m^2_t/2}{\hat s+m^2_t/2}\right),
\end{eqnarray}
where
\begin{eqnarray}
\hat\sigma(\cos\theta_\ell >0)=\int^1_0d\cos\theta_\ell \frac{d\hat\sigma}{d\cos\theta_\ell }~~{\rm and}~~\hat\sigma(\cos\theta_\ell <0)=\int^0_{-1}d\cos\theta_\ell \frac{d\hat\sigma}{d\cos\theta_\ell }
\end{eqnarray}
Assuming the $\wpri$ is on mass shell, the hadronic forward-backward asymmetry is the same as the partonic forward-backward asymmetry with $\hat s=M^2_{W'}$.

%%%%%%%%%%%%%%%%%%%%%%%%%%%%%%%%%%%%%%%%%%%%%%%%%%%%%%%%%%%%%%%
\section{EVALUATION OF ${\cal I}_1$}
\label{ContInt.APP}
Here we give the details of how to integrate 
\begin{eqnarray}
\mathcal{I}_1(\cos\theta_\ell )=\frac{1}{\pi^2}\int^{1}_{-1}d\cos\theta_3\mathcal{I}_{\phi_-}(\cos\theta_3,\cos\theta_\ell )
\end{eqnarray}
where
\begin{eqnarray}
\mathcal{I}_{\phi_-}(\cos\theta_3,\cos\theta_\ell )=\int^\pi_0d\phi_-(\pi-\phi_-)\frac{1+\cos\theta_{3l}}{[1+\cos\theta_{3l}+x^2_{Wt}(1-\cos\theta_{3l})]^2}
\label{eq:Iphi}
\end{eqnarray}
First, through a change of variable, $\phi'=2(\pi-\phi_-)$, we can rewrite $\mathcal{I}_{\phi_-}$ as
\begin{eqnarray}
\mathcal{I}_{\phi_-}(\cos\theta_3,\cos\theta_\ell )=\frac{1}{4}\int^{2\pi}_0d\phi' \phi'\frac{C+B\cos\phi'}{(F+D\cos\phi')^2},
\end{eqnarray}
where 
\begin{eqnarray}
C=1+\cos\theta_\ell \cos\theta_3,&&B=\sin\theta_3\sin\theta_\ell ,\nonumber\\
F=1+x^2_{Wt}+(1-x^2_{Wt})\cos\theta_\ell \cos\theta_3, &{\rm and}& D=(1-x^2_{Wt})\sin\theta_\ell \sin\theta_3.\label{eq:ABDF}
\end{eqnarray} 

\begin{figure}[t]
\begin{center}
\includegraphics[width=0.47\textwidth,clip=true]{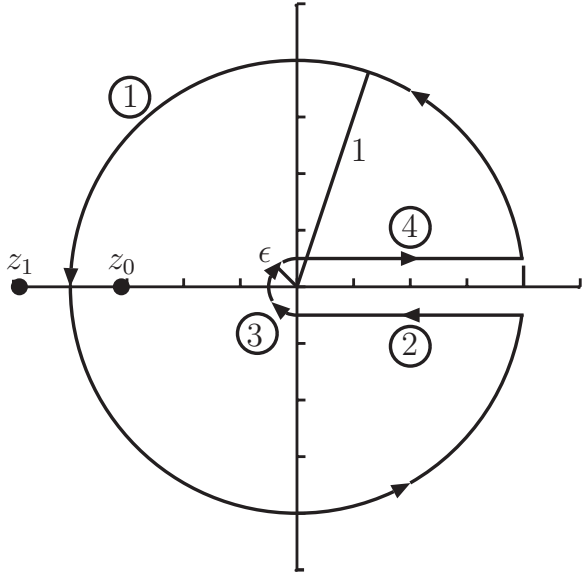}

\caption{Contour integral for the lepton angular distribution calculation.
The points $z_0$ and $z_1$ are the poles of the integrand in Eq.~(\ref{eq:contour}). }
 \label{fig:contour}
\end{center}
\end{figure}

To compute $I_{\phi_-}$,we will consider the contour integral
\begin{eqnarray}
I_C=\oint_C dz \log z \frac{2Cz+B(z^2+1)}{(Dz^2+2Fz+D)^2},
\label{eq:Ic}
\label{eq:contour}
\end{eqnarray}
where the contour $C$ is illustrated in Fig.~\ref{fig:contour}.  We have defined the branch cut of the complex logarithm to be along the positive real axis and deformed the contour about this branch cut.  The contour is then a closed loop that consists of these segments:
\begin{description}
\setlength{\itemindent}{5pt}
\item[Segment 1:] an arc of a unit circle running for $1+i\epsilon$ to $1-i\epsilon$
\item[Segment 2:] a straight line from $1-i\epsilon$ to $-i\epsilon$
\item[Segment 3:] half circle of radius $\epsilon$ centered about zero running from $-i\epsilon$ to $i\epsilon$
\item[Segment 4:] a straight line from $i\epsilon$ to $1+i\epsilon$
\end{description}
As we will show, segment 1 is proportional to the integral $I_{\phi_-}$.

Using Cauchy's integral formula we can relate the contour integral $I_C$ to the residues of the integrand at the integrand's poles.  The integrand in Eq.~(\ref{eq:Ic}) has two poles at
\begin{eqnarray}
z_0=\frac{-F+\sqrt{F^2-D^2}}{D}~~{\rm and}~~z_1=\frac{-F-\sqrt{F^2-D^2}}{D}.
\end{eqnarray}
Since $F>D$, the pole $z_0$ ($z_1$) lies on the negative real axis inside (outside) the unit circle centered about zero.  These poles are shown in Fig.~\ref{fig:contour}.  As can be seen, $z_0$ is the only pole inside the contour $C$, and we can use the Cauchy integral formula to find
\begin{eqnarray}
I_C&=&2\pi i~\text{Res}_{z=z_0}~\log z\frac{2Cz+B(z^2+1)}{(Dz^2+2Fz+D)^2}\nonumber\\
&=&i\pi\frac{(CD-BF)\sqrt{F^2-D^2}-D(BD-CF)\log(z_0)}{D(F^2-D^2)^{3/2}}
\label{eq:residue}
\end{eqnarray}

Now the contour will be integrated along all four segments and the limit $\epsilon\rightarrow 0$ will be taken.  Segment 1 is parametrized by  $z=e^{i\phi}$ and $\phi$ is integrated from $\epsilon$ to $2\pi-\epsilon$:
\begin{eqnarray}
I_1&=&-\lim_{\epsilon\rightarrow 0}\frac{1}{2}\int^{2\pi-\epsilon}_{\epsilon}d\phi\frac{C+B\cos\phi}{(F+D\cos\phi)^2}\nonumber\\
&=&-2I_{\phi_-}
\end{eqnarray}
As mentioned earlier, $I_1$ is proportional to $I_{\phi_-}$.

\par
The parametrization of segment 2 is $z=x-i\epsilon$ and $x$ is integrated from $1$ to $0$:
\begin{eqnarray}
I_2&=&\lim_{\epsilon\rightarrow0}\int^{0}_{1}dx\log(x-i\epsilon)\frac{2C(x-i\epsilon)+B((x-i\epsilon)^2+1)}{(2F(x-i\epsilon)+D((x-i\epsilon)^2+1))^2}
\end{eqnarray}
Since limits of the integral do not depend on $\epsilon$ and the integrand is well behaved as $\epsilon\rightarrow 0$, we can compute the $\epsilon\rightarrow 0$ limit of the integrand and then integrate.  The limit of $\log(x-i\epsilon)$ as $\epsilon$ goes to zero must be taken carefully.   Since the limit as $\epsilon\rightarrow 0$ approaches the positive real axis from below, we have
\begin{eqnarray}
\lim_{\epsilon\rightarrow 0}\log(x-i\epsilon)=\log |x|+i\lim_{\epsilon\rightarrow 0}\arctan\frac{-\epsilon}{x}=\log x+2i\pi,
\end{eqnarray}
Then $I_2$ becomes
\begin{eqnarray}
I_2&=&-\int^{1}_{0}dx\log x\frac{2Cx+B(x^2+1)}{(2Fx+D(x^2+1))^2}-2i\pi\int^{1}_{0}dx\frac{2Cx+B(x^2+1)}{(2Fx+D(x^2+1))^2}.
\label{eq:I2}
\end{eqnarray}
The second integral in Eq.~(\ref{eq:I2}) can be performed analytically to find
\begin{eqnarray}
-2i\pi\int^{1}_{0}dx\frac{2Cx+B(x^2+1)}{(2Fx+D(x^2+1))^2}&=&i\pi\frac{(CD-BF)\sqrt{F^2-D^2}-D(BD-CF)\log(-z_0)}{D(F^2-D^2)^{3/2}}\nonumber\\
&=&I_C+\pi^2\frac{CF-BD}{(F^2-D^2)^{3/2}},
\end{eqnarray}
where Eq.~(\ref{eq:residue}) has been used to express the integral in terms of $I_C$.
Finally, the integral along segment 2 is
\begin{eqnarray}
I_2&=&-\int^{1}_{0}dx\log x\frac{2Cx+B(x^2+1)}{(2Fx+D(x^2+1))^2}+I_C+\pi^2\frac{CF-BD}{(F^2-D^2)^{3/2}}
\end{eqnarray}

\par
For segment 3 we set $z=\epsilon e^{i\phi}$ and integrate $\phi$ from $3\pi/2$ to $\pi/2$.  The integral is then
\begin{eqnarray}
I_3&=&\lim_{\epsilon\rightarrow 0}\bigg{\{}-\epsilon \int^{\pi/2}_{3\pi/2}d\phi e^{i\phi}\phi\frac{2C\epsilon e^{i\phi}+B(\epsilon^2e^{2i\phi}+1)}{(D\epsilon^2e^{2i\phi}+2F\epsilon e^{i\phi}+D)^2}\nonumber\\
&&+i\epsilon\log\epsilon\int^{\pi/2}_{3\pi/2}d\phi e^{i\phi}\frac{2C\epsilon e^{i\phi}+B(\epsilon^2e^{2i\phi}+1)}{(D\epsilon^2e^{2i\phi}+2F\epsilon e^{i\phi}+D)^2}\bigg{\}},
\label{eq:I3}
\end{eqnarray}
As $\epsilon$ goes to zero, both the integrals in Eq.~(\ref{eq:I3}) are finite and their coefficients go to zero.  Hence, %in the limit as $\epsilon$ goes to zero,
\begin{eqnarray}
I_3 = 0
\end{eqnarray}
\par
Segment 4 is the final segment to be integrated.  This segment is parametrized by $z=x+i\epsilon$ and $x$ is integrated from $0$ to $1$.  The integral is then
\begin{eqnarray}
I_4&=&\lim_{\epsilon\rightarrow 0}\int^{1}_{0}dx\log(x+i\epsilon)\frac{2C(x+i\epsilon)+B((x+i\epsilon)^2+1)}{(2F(x+i\epsilon)+D((x+i\epsilon)^2+1))^2}\nonumber\\
&=&\int^{1}_{0}dx\log x\frac{2Cx+B(x^2+1)}{(2Fx+D(x^2+1))^2}.
\end{eqnarray}
As discussed for segment 2, the integral and $\epsilon$ limit can be computed separately since the limits of the integral do not depend on $\epsilon$ and the integrand is well behaved in the $\epsilon\rightarrow 0$ limit.  Also, unlike segment 2, the positive real axis is approached from above.  Hence, the $\log$ does not pick up the angle $2i\pi$ that was found for $I_2$.

Now we are ready to calculate the total contour integral.  This is accomplished by adding all four segments together:
\begin{eqnarray}
I_C=I_1+I_2+I_3+I_4=I_C+\pi^2\frac{CF-BD}{(F^2-D^2)^{3/2}}-2I_{\phi_-}.
\end{eqnarray}
Hence, $I_{\phi_-}$ is found to be
\begin{eqnarray}
I_{\phi_-}=\frac{\pi^2}{2}\frac{CF-BD}{(F^2-D^2)^{3/2}}
\end{eqnarray}

Finally, using the definitions of $C,B,F$, and $D$ in Eq.~(\ref{eq:ABDF}) and integrating over $\cos\theta_3$, we can compute
\begin{eqnarray}
\mathcal{I}_1(\cos\theta_\ell )=\frac{x^2_{wt}-1-\log x^2_{wt}}{2(1-x^2_{wt})^2}.
\end{eqnarray}

%%%%%%%%%%%%%%%%%%%%%%%%%%%%%%%%%%%%%%%%%%%%%%%%%%%%%%%%%%%%%%%%%%%%%%%%%%%%%%%%%

\pagebreak

%%%%%%%%%%%%%%%%%%%%%%%%%%%%%%%%%%%%%%%%%%%%%%%%%%%%%%%%%%%%%%%%%%%%%%%%%%%%%%%%%%


\begin{thebibliography}{99}

\bibitem{ATLASTDR}
ATLAS Collaboration, Detector and Physics Performances Technical Design
Report, Vol.II, ATLAS TDR 14, CERN/LHCC 99-14,
  
\bibitem{CMSTDR}
CMS Physics TDR, Volume II: CERN-LHCC-2006-021, 25 June 2006
As published in J. Phys. G: Nucl. Part. Phys. 34 995-1579 (2006)   

%\cite{Pati:1974yy}
\bibitem{Pati:1974yy}
  J.~C.~Pati and A.~Salam,
  %``Lepton Number As The Fourth Color,''
  Phys.\ Rev.\  D {\bf 10}, 275 (1974)
  [Erratum-ibid.\  D {\bf 11}, 703 (1975)]; 
  %%CITATION = PHRVA,D10,275;%%
  %\cite{Senjanovic:1975rk}
%\bibitem{Senjanovic:1975rk}
  G.~Senjanovic and R.~N.~Mohapatra,
  %``Exact Left-Right Symmetry And Spontaneous Violation Of Parity,''
  Phys.\ Rev.\  D {\bf 12}, 1502 (1975).
  %%CITATION = PHRVA,D12,1502;%%

%\cite{Langacker:1984dc}
\bibitem{Langacker:1984dc}
  P.~Langacker, R.~W.~Robinett and J.~L.~Rosner,
  %``New Heavy Gauge Bosons In P P And P Anti-P Collisions,''
  Phys.\ Rev.\  D {\bf 30}, 1470 (1984); 
  %%CITATION = PHRVA,D30,1470;%%
%
%\cite{Dittmar:2003ir}
%\bibitem{Dittmar:2003ir}
  M.~Dittmar, A.~S.~Nicollerat and A.~Djouadi,
  %``Z' studies at the LHC: An update,''
  Phys.\ Lett.\  B {\bf 583}, 111 (2004)
  [arXiv:hep-ph/0307020]; 
  %%CITATION = PHLTA,B583,111;%%
%\cite{Carena:2004xs}
%\bibitem{Carena:2004xs}
  M.~S.~Carena, A.~Daleo, B.~A.~Dobrescu and T.~M.~P.~Tait,
  %``$Z^\prime$ gauge bosons at the Tevatron,''
  Phys.\ Rev.\  D {\bf 70}, 093009 (2004)
  [arXiv:hep-ph/0408098]; 
  %%CITATION = PHRVA,D70,093009;%%
%\cite{Han:2005ru}
%\bibitem{Han:2005ru}
  T.~Han, H.~E.~Logan and L.~T.~Wang,
  %``Smoking-gun signatures of little Higgs models,''
  JHEP {\bf 0601}, 099 (2006)
  [arXiv:hep-ph/0506313];
  %%CITATION = JHEPA,0601,099;%%
%\cite{Petriello:2008zr}
%\bibitem{Petriello:2008zr}
  F.~Petriello and S.~Quackenbush,
  %``Measuring $Z^\prime$ couplings at the CERN LHC,''
  Phys.\ Rev.\  D {\bf 77}, 115004 (2008)
  [arXiv:0801.4389 [hep-ph]].
  %%CITATION = PHRVA,D77,115004;%%

%\cite{Cvetic:1993ska}
\bibitem{Cvetic:1993ska}
  M.~Cvetic, P.~Langacker and J.~Liu,
  %``Testing the handedness of a heavy W-prime at future hadron colliders,''
  Phys.\ Rev.\  D {\bf 49}, 2405 (1994)
  [arXiv:hep-ph/9308251].
  %%CITATION = PHRVA,D49,2405;%%

%\cite{Agashe:2007ki}
\bibitem{Agashe:2007ki}
  K.~Agashe {\it et al.},
  %``LHC Signals for Warped Electroweak Neutral Gauge Bosons,''
  Phys.\ Rev.\  D {\bf 76}, 115015 (2007)
  [arXiv:0709.0007 [hep-ph]].
  %%CITATION = PHRVA,D76,115015;%%

%\cite{Agashe:2008jb}
\bibitem{Agashe:2008jb}
  K.~Agashe, S.~Gopalakrishna, T.~Han, G.~Y.~Huang and A.~Soni,
  %``LHC Signals for Warped Electroweak Charged Gauge Bosons,''
  Phys.\ Rev.\  D {\bf 80}, 075007 (2009)
  [arXiv:0810.1497 [hep-ph]].
  %%CITATION = PHRVA,D80,075007;%%


%\cite{Djouadi:2007eg}
\bibitem{Djouadi:2007eg}
%\cite{Agashe:2006hk}
%\bibitem{Agashe:2006hk}
  K.~Agashe, A.~Belyaev, T.~Krupovnickas, G.~Perez and J.~Virzi,
  %``LHC signals from warped extra dimensions,''
  Phys.\ Rev.\  D {\bf 77}, 015003 (2008)
  [arXiv:hep-ph/0612015]; 
  %%CITATION = PHRVA,D77,015003;%%
  A.~Djouadi, G.~Moreau and R.~K.~Singh,
  %``Kaluza--Klein excitations of gauge bosons at the LHC,''
  Nucl.\ Phys.\  B {\bf 797}, 1 (2008)
  [arXiv:0706.4191 [hep-ph]].
  %%CITATION = NUPHA,B797,1;%%

%\cite{Boos:2006xe}
\bibitem{Boos:2006xe}
  E.~Boos, V.~Bunichev, L.~Dudko and M.~Perfilov,
  %``Interference between $W^\prime$ and $W$ in single-top-quark production
  %processes,''
  Phys.\ Lett.\  B {\bf 655}, 245 (2007)
  [arXiv:hep-ph/0610080].
  %%CITATION = PHLTA,B655,245;%%


%\cite{Frere:1990qm}
\bibitem{Frere:1990qm}
  J.~M.~Frere and W.~W.~Repko,
  %``W(R) identification at hadron colliders,''
  Phys.\ Lett.\  B {\bf 254}, 485 (1991).
  %%CITATION = PHLTA,B254,485;%%

\bibitem{InsenWtb}
%\cite{Grzadkowski:1999iq}
%\bibitem{Grzadkowski:1999iq}
  B.~Grzadkowski and Z.~Hioki,
  %``New hints for testing anomalous top quark interactions at future linear
  %colliders,''
  Phys.\ Lett.\  B {\bf 476}, 87 (2000)
  [arXiv:hep-ph/9911505];
  %%CITATION = PHLTA,B476,87;%%
%
%\cite{Rindani:2000jg}
%\bibitem{Rindani:2000jg}
  S.~D.~Rindani,
  %``Effect of anomalous t b W vertex on decay-lepton distributions in  e+ e-
  %--> t anti-t and CP-violating asymmetries,''
  Pramana {\bf 54}, 791 (2000)
  [arXiv:hep-ph/0002006];
  %%CITATION = PRAMC,54,791;%%
%
%\cite{Grzadkowski:2002gt}
%\bibitem{Grzadkowski:2002gt}
  B.~Grzadkowski and Z.~Hioki,
  %``Decoupling of anomalous top-decay vertices in angular distribution of
  %secondary particles,''
  Phys.\ Lett.\  B {\bf 557}, 55 (2003)
  [arXiv:hep-ph/0208079];
  %%CITATION = PHLTA,B557,55;%%
%
%\cite{Godbole:2006tq}
%\bibitem{Godbole:2006tq}
  R.~M.~Godbole, S.~D.~Rindani and R.~K.~Singh,
  %``Lepton distribution as a probe of new physics in production and decay of
  %the t quark and its polarization,''
  JHEP {\bf 0612}, 021 (2006)
  [arXiv:hep-ph/0605100];
  %%CITATION = JHEPA,0612,021;%%
%
%\cite{Godbole:2009dp}
%\bibitem{Godbole:2009dp}
  R.~M.~Godbole, S.~D.~Rindani, K.~Rao and R.~K.~Singh,
  %``Top polarization as a probe of new physics,''
  AIP Conf.\ Proc.\  {\bf 1200}, 682 (2010)
  [arXiv:0911.3622 [hep-ph]].
  %%CITATION = APCPC,1200,682;%%

%\cite{ArkaniHamed:2002qy}
\bibitem{ArkaniHamed:2002qy}
  N.~Arkani-Hamed, A.~G.~Cohen, E.~Katz and A.~E.~Nelson,
  %``The littlest Higgs,''
  JHEP {\bf 0207}, 034 (2002)
  [arXiv:hep-ph/0206021]; 
  %%CITATION = JHEPA,0207,034;%%
%\cite{Kaplan:2003uc}
%\bibitem{Kaplan:2003uc}
  D.~E.~Kaplan and M.~Schmaltz,
  %``The little Higgs from a simple group,''
  JHEP {\bf 0310}, 039 (2003)
  [arXiv:hep-ph/0302049];
  %%CITATION = JHEPA,0310,039;%%
%\cite{Han:2003wu}
%\bibitem{Han:2003wu}
  T.~Han, H.~E.~Logan, B.~McElrath and L.~T.~Wang,
  %``Phenomenology of the little Higgs model,''
  Phys.\ Rev.\  D {\bf 67}, 095004 (2003)
  [arXiv:hep-ph/0301040].
  %%CITATION = PHRVA,D67,095004;%%

\bibitem{Appelquist:2000nn}
  T.~Appelquist, H.~C.~Cheng and B.~A.~Dobrescu,
  %``Bounds on universal extra dimensions,''
  Phys.\ Rev.\  D {\bf 64}, 035002 (2001)
  [arXiv:hep-ph/0012100]; 
  %%CITATION = PHRVA,D64,035002;%%
%\cite{Cheng:2002ab}
%\bibitem{Cheng:2002ab}
  H.~C.~Cheng, K.~T.~Matchev and M.~Schmaltz,
  %``Bosonic supersymmetry? Getting fooled at the CERN LHC,''
  Phys.\ Rev.\  D {\bf 66}, 056006 (2002)
  [arXiv:hep-ph/0205314].
  %%CITATION = PHRVA,D66,056006;%%

%\cite{Csaki:2003zu}
\bibitem{Csaki:2003zu}
  C.~Csaki, C.~Grojean, L.~Pilo and J.~Terning,
  %``Towards a realistic model of Higgsless electroweak symmetry breaking,''
  Phys.\ Rev.\ Lett.\  {\bf 92}, 101802 (2004)
  [arXiv:hep-ph/0308038];
  %%CITATION = PRLTA,92,101802;%%
%\cite{Chivukula:2006cg}
%\bibitem{Chivukula:2006cg}
  R.~S.~Chivukula, B.~Coleppa, S.~Di Chiara, E.~H.~Simmons, H.~J.~He, M.~Kurachi and M.~Tanabashi,
  %``A three site higgsless model,''
  Phys.\ Rev.\  D {\bf 74}, 075011 (2006)
  [arXiv:hep-ph/0607124];
  %%CITATION = PHRVA,D74,075011;%%
%\cite{He:2007ge}
%\bibitem{He:2007ge}
  H.~J.~He {\it et al.},
  %``LHC Signatures of New Gauge Bosons in Minimal Higgsless Model,''
  Phys.\ Rev.\  D {\bf 78}, 031701 (2008)
  [arXiv:0708.2588 [hep-ph]].
  %%CITATION = PHRVA,D78,031701;%%

%\cite{Agashe:2003zs}
\bibitem{Agashe:2003zs}
  K.~Agashe, A.~Delgado, M.~J.~May and R.~Sundrum,
  %``RS1, custodial isospin and precision tests,''
  JHEP {\bf 0308}, 050 (2003)
  [arXiv:hep-ph/0308036].
  %%CITATION = JHEPA,0308,050;%%

\bibitem{Abulencia:2006kh}
  A.~Abulencia {\it et al.}  [CDF Collaboration],
  %``Search for W' boson decaying to electron-neutrino pairs in p anti-p
  %collisions at s**(1/2) = 1.96-TeV,''
  Phys.\ Rev.\  D {\bf 75}, 091101 (2007)
  [arXiv:hep-ex/0611022].
  %%CITATION = PHRVA,D75,091101;%%
%\bibitem{CDF-8747}
%	CDF Collaboration, note-8747(2007).

\bibitem{:2007bs}
  V.~M.~Abazov {\it et al.}  [D0 Collaboration],
  %``Search for W' bosons decaying to an electron and a neutrino with the D0
  %detector,''
  Phys.\ Rev.\ Lett.\  {\bf 100}, 031804 (2008)
  [arXiv:0710.2966 [hep-ex]].
  %%CITATION = PRLTA,100,031804;%%


\bibitem{Aaltonen:2009qu}
  T.~Aaltonen {\it et al.}  [CDF Collaboration],
  %``Search for the Production of Narrow tb Resonances in 1.9 fb-1 of ppbar
  %Collisions at sqrt(s) = 1.96 TeV,''
  Phys.\ Rev.\ Lett.\  {\bf 103}, 041801 (2009)
  [arXiv:0902.3276 [hep-ex]].
  %%CITATION = PRLTA,103,041801;%%

%\cite{Hsieh:2010zr}
\bibitem{Hsieh:2010zr}
  K.~Hsieh, K.~Schmitz, J.~H.~Yu and C.~P.~Yuan,
  %``Global Analysis of General SU(2) x SU(2) x U(1) Models with Precision
  %Data,''
  Phys.\ Rev.\  D {\bf 82}, 035011 (2010)
  [arXiv:1003.3482 [hep-ph]].
  %%CITATION = PHRVA,D82,035011;%%


%\cite{Cacciapaglia:2006pk}
\bibitem{Cacciapaglia:2006pk}
  G.~Cacciapaglia, C.~Csaki, G.~Marandella and A.~Strumia,
  %``The minimal set of electroweak precision parameters,''
  Phys.\ Rev.\  D {\bf 74}, 033011 (2006)
  [arXiv:hep-ph/0604111].
  %%CITATION = PHRVA,D74,033011;%%

\bibitem{Czakon:1999ga}
  M.~Czakon, J.~Gluza and M.~Zralek,
  %``Low energy physics and left-right symmetry: Bounds on the model
  %parameters,''
  Phys.\ Lett.\  B {\bf 458}, 355 (1999)
  [arXiv:hep-ph/9904216].
  %%CITATION = PHLTA,B458,355;%%

\bibitem{LR-bound}
%\bibitem{Beall:1981ze}
  G.~Beall, M.~Bander and A.~Soni,
  %``Constraint On The Mass Scale Of A Left-Right Symmetric Electroweak Theory
  %From The K(L) K(S) Mass Difference,''
  Phys.\ Rev.\ Lett.\  {\bf 48}, 848 (1982);
  %%CITATION = PRLTA,48,848;%%
%
%\bibitem{Ecker:1985vv}
  G.~Ecker and W.~Grimus,
  %``CP Violation And Left-Right Symmetry,''
  Nucl.\ Phys.\  B {\bf 258}, 328 (1985);
  %%CITATION = NUPHA,B258,328;%%
%
%\bibitem{Zhang:2007fn}
  Y.~Zhang, H.~An, X.~Ji and R.~N.~Mohapatra,
  %``Right-handed quark mixings in minimal left-right symmetric model with
  %general CP violation,''
  Phys.\ Rev.\  D {\bf 76}, 091301 (2007)
  [arXiv:0704.1662 [hep-ph]];
  %%CITATION = PHRVA,D76,091301;%%
%
%\cite{Maiezza:2010ic}
%\bibitem{Maiezza:2010ic}
  A.~Maiezza, M.~Nemevsek, F.~Nesti and G.~Senjanovic,
  %``Left-Right Symmetry at LHC,''
  Phys.\ Rev.\  D {\bf 82}, 055022 (2010)
  [arXiv:1005.5160 [hep-ph]].
  %%CITATION = PHRVA,D82,055022;%%

%\cite{Langacker:1989xa}
\bibitem{Langacker:1989xa}
  P.~Langacker and S.~U.~Sankar,
  %``Bounds on the Mass of W(R) and the W(L)-W(R) Mixing Angle xi in General
  %SU(2)-L x SU(2)-R x U(1) Models,''
  Phys.\ Rev.\  D {\bf 40}, 1569 (1989).
  %%CITATION = PHRVA,D40,1569;%%

% mordified by Yu-Feng, v5i
\bibitem{Wu:2007kt}
  Y.~L.~Wu and Y.~F.~Zhou,
  %``Two Higgs Bi-doublet Left-Right Model With Spontaneous P and CP
  %Violation,''
  Sci.\ China {\bf G51}, 1808 (2008)
  [arXiv:0709.0042 [hep-ph]];
  %%CITATION = SCASE,G51,1808;%%
%\cite{Wu:2007gb}
%\bibitem{Wu:2007gb}
%  Y.~L.~Wu and Y.~F.~Zhou,
  %``A Two Higgs Bi-doublet Left-Right Model With Spontaneous CP Violation,''
  Int.\ J.\ Mod.\ Phys.\  A {\bf 23}, 3304 (2008)
  [arXiv:0711.3891 [hep-ph]];
  %%CITATION = IMPAE,A23,3304;%%
%

\bibitem{Guadagnoli:2010sd}
  D.~Guadagnoli and R.~N.~Mohapatra,
  %``TeV Scale Left Right Symmetry and Flavor Changing Neutral Higgs Effects,''
  arXiv:1008.1074 [hep-ph].
  %%CITATION = ARXIV:1008.1074;%%

%\cite{Ball:2007zza}
\bibitem{Ball:2007zza}
  G.~L.~Bayatian {\it et al.}  [CMS Collaboration],
  %``CMS technical design report, volume II: Physics performance,''
  J.\ Phys.\ G {\bf 34}, 995 (2007);
  %%CITATION = JPHGB,G34,995;%%
%\cite{Aad:2009wy}
%\bibitem{Aad:2009wy}
  G.~Aad {\it et al.}  [The ATLAS Collaboration],
  %``Expected Performance of the ATLAS Experiment - Detector, Trigger and
  %Physics,''
  arXiv:0901.0512 [hep-ex].
  %%CITATION = ARXIV:0901.0512;%%

%\cite{Rizzo:2007xs}
\bibitem{Rizzo:2007xs}
  T.~G.~Rizzo,
  %``The Determination of the Helicity of $W'$ Boson Couplings at the LHC,''
  JHEP {\bf 0705}, 037 (2007)
  [arXiv:0704.0235 [hep-ph]].
  %%CITATION = JHEPA,0705,037;%%

%\cite{Barger:2006hm}
\bibitem{Barger:2006hm}
  V.~Barger, T.~Han and D.~G.~E.~Walker,
  %``Top Quark Pairs at High Invariant Mass - A Model-Independent Discriminator of New Physics at the LHC,''
  Phys.\ Rev.\ Lett.  {\bf 100}, 031801 (2008)
  [arXiv:hep-ph/0612016].

%\cite{Kaplan:2008tt}
\bibitem{Kaplan:2008tt}
  D.~E.~Kaplan, K.~Rehermann, M.~D.~Schwartz and B.~Tweedie, 
  %``Top-tagging: A Method for Identifying Boosted Hadronic Tops,''
  Phys.\ Rev.\ Lett.  {\bf 101}, 142001 (2008)
  [arXiv:0806.0848v2[hep-ph]]; 
%%CITATION = PRLTA,101,142001;%%
%  
%\cite{Thaler:2008w}
%\bibitem{Thaler:2008w}
  J.~Thaler and L.-T.~Wang,
  %``Strategies to Identify Boosted Tops,''
  JHEP  {\bf 0807}, 092 (2008)
  [arXiv:0806.0023v2[hep-ph]].
  %%CITATION = JHEPA,0806,092;%%

%\cite{Almeida:2008yp}
\bibitem{Almeida:2008yp}
  L.~G.~Almeida, S.~J.~Lee, G.~Perez, G.~F.~Sterman, I.~Sung and J.~Virzi,
  %``Substructure of high-$p_T$ Jets at the LHC,''
  Phys.\ Rev.\  D {\bf 79}, 074017 (2009)
  [arXiv:0807.0234 [hep-ph]]; 
  %%CITATION = PHRVA,D79,074017;%%
%\cite{Almeida:2008tp}
%\bibitem{Almeida:2008tp}
  L.~G.~Almeida, S.~J.~Lee, G.~Perez, I.~Sung and J.~Virzi,
  %``Top Jets at the LHC,''
  Phys.\ Rev.\  D {\bf 79}, 074012 (2009)
  [arXiv:0810.0934 [hep-ph]].
  %%CITATION = PHRVA,D79,074012;%%

%\cite{Dittmar:1996my}
\bibitem{Dittmar:1996my}
  M.~Dittmar,
  %``Neutral current interference in the TeV region: The experimental
  %sensitivity at the LHC,''
  Phys.\ Rev.\  D {\bf 55}, 161 (1997)
  [arXiv:hep-ex/9606002].
  %%CITATION = PHRVA,D55,161;%%

%\cite{Tait:2000sh}
\bibitem{Tait:2000sh}
  T.~M.~P.~Tait and C.~P.~P.~Yuan,
  %``Single top quark production as a window to physics beyond the standard
  %model,''
  Phys.\ Rev.\  D {\bf 63}, 014018 (2000)
  [arXiv:hep-ph/0007298].
  %%CITATION = PHRVA,D63,014018;%%

%\cite{Bernreuther:2001rq}
\bibitem{Bernreuther:2001rq}
  W.~Bernreuther, A.~Brandenburg, Z.~G.~Si and P.~Uwer,
  %``Top quark spin correlations at hadron colliders: Predictions at
  %next-to-leading order QCD,''
  Phys.\ Rev.\ Lett.\  {\bf 87}, 242002 (2001)
  [arXiv:hep-ph/0107086];
  %%CITATION = PRLTA,87,242002;%%
%\cite{Bernreuther:2004jv}
%\bibitem{Bernreuther:2004jv}
%  W.~Bernreuther, A.~Brandenburg, Z.~G.~Si and P.~Uwer,
  %``Top quark pair production and decay at hadron colliders,''
  Nucl.\ Phys.\  B {\bf 690}, 81 (2004)
  [arXiv:hep-ph/0403035];
  %%CITATION = NUPHA,B690,81;%%
%\cite{Bernreuther:2010ny}
%\bibitem{Bernreuther:2010ny}
  W.~Bernreuther and Z.~G.~Si,
  %``Distributions and correlations for top quark pair production and decay at
  %the Tevatron and LHC,''
  Nucl.\ Phys.\  B {\bf 837}, 90 (2010)
  [arXiv:1003.3926 [hep-ph]].
  %%CITATION = NUPHA,B837,90;%%


%\cite{Acosta:2004wq}
%\bibitem{Acosta:2004wq}
%  D.~E.~Acosta {\it et al.}  [CDF Collaboration],
  %``Measurement of the forward-backward charge asymmetry of electron positron
  %pairs in $p\bar{p}$ collisions at $\sqrt{s} = 1.96$ TeV,''
%  Phys.\ Rev.\  D {\bf 71}, 052002 (2005)
%  [arXiv:hep-ex/0411059].
  %%CITATION = PHRVA,D71,052002;%%

\bibitem{OtherTopPol}
%\cite{Shelton:2008nq}
%\bibitem{Shelton:2008nq}
  J.~Shelton,
  %``Polarized tops from new physics: signals and observables,''
  Phys.\ Rev.\  D {\bf 79}, 014032 (2009)
  [arXiv:0811.0569 [hep-ph]];
  %%CITATION = PHRVA,D79,014032;%%
%\cite{Krohn:2009wm}
%\bibitem{Krohn:2009wm}
  D.~Krohn, J.~Shelton and L.~T.~Wang,
  %``Measuring the Polarization of Boosted Hadronic Tops,''
  JHEP {\bf 1007}, 041 (2010)
  [arXiv:0909.3855 [hep-ph]].
  %%CITATION = JHEPA,1007,041;%%

\end{thebibliography}
\end{document}